\documentclass{aa}  

\usepackage{graphicx}
\usepackage{subcaption}
\usepackage{comment}
\usepackage{txfonts}
\usepackage{hyperref}
\usepackage{color}
\usepackage{booktabs,caption}
\usepackage{longtable}
\usepackage[flushleft]{threeparttable}
\usepackage{cuted}
\usepackage{multicol}
\newcommand{\ii}{\,{\sc ii}}
\newcommand{\iii}{\,{\sc iii}}

\begin{document}

   \title{Optical Emission-Line Properties of \\ eROSITA-selected SDSS-V Galaxies}

   \titlerunning{eROSITA SDSS-V selected galaxies}

   \authorrunning{N. G. Pulatova, E. Rubtsov et al.}
\author{Nadiia G. Pulatova
          \inst{1} \fnmsep\thanks{These authors contributed equally to this work.}
          \and
          Evgenii Rubtsov\inst{2,\star}
          \and
          Igor V. Chilingarian\inst{3,2}
          \and
          Hans-Walter Rix \inst{1}
          \and
          Mariia Demianenko \inst{1,4}
          \and
          Kirill~A.~Grishin \inst{5, 2}
          \and
          Ivan Yu. Katkov \inst{6, 2}
          \and
          Donald P. Schneider \inst{7, 18}
          \and
          Catarina Aydar \inst{8, 9}
          \and
          Johannes Buchner \inst{8}
          \and
          Mara~Salvato \inst{8}
          \and
          Andrea Merloni \inst{8}
          \and
          Anton M. Koekemoer \inst{10}
          \and
          Roberto J. Assef \inst{11}
          \and 
          Claudio Ricci \inst{11}
          \and
          Dominika Wylezalek \inst{12}
          \and
          Damir Gasymov \inst{12, 2, 4}
          \and
           William Nielsen Brandt \inst{13}
          \and
          Castalia Alenka Negrete Peñaloza \inst{14}
          \and
          Sean Morrison \inst{15}
          \and
          Scott~F.~Anderson \inst{16}
          \and
          Franz~E.~Bauer \inst{17}
          \and
          Hector Javier Ibarra-Medel \inst{14}
          \and
          Qiaoya Wu \inst{15}
     }

        \institute{Max-Planck-Institute for Astronomy, K\"onigstuhl 17, D-69117     Heidelberg, Germany\\
              \email{pulatova@mpia.de}
        \and   
            Sternberg Astronomical Institute, M.V. Lomonosov Moscow State University, Universitetskiy prosp. 13, Moscow, 119234, Russia 
        \and
            Center for Astrophysics | Harvard \& Smithsonian, 60 Garden St. MS09, Cambridge, MA 02138, USA 
        \and
            Department for Physics and Astronomy, Heidelberg University, Im Neuenheimer Feld 226, 69120 Heidelberg, Germany
        \and
            Universite Paris Cite, CNRS, Astroparticule et Cosmologie, F-75013 Paris, France
        \and
             New York University Abu Dhabi, PO Box 129188, Abu Dhabi, UAE 
        \and
             Department of Astronomy \& Astrophysics, The Pennsylvania State University, University Park, PA 16802, USA 
        \and
             Max-Planck-Institut f\"ur Extraterrestrische Physik, Giessenbachstrasse, 85748 Garching, Germany 
        \and
            Exzellenzcluster ORIGINS, Boltzmannstr. 2, 85748 Garching, Germany
        \and
            Space Telescope Science Institute, 3700 San Martin Drive, Baltimore, MD 21218, USA
        \and
            Instituto de Estudios Astrof\'isicos, Facultad de Ingenier\'ia y Ciencias, Universidad Diego Portales, Av. Ej\'ercito Libertador 441, Santiago, Chile
        \and
            Astronomisches Rechen-Institut, Zentrum f\"ur Astronomie der Universit\"at Heidelberg, M\"onchhofstr. 12-14, D-69120 Heidelberg, Germany 
        \and
            Department of Astronomy \& Astrophysics, The Pennsylvania State University, University Park, PA 16802, USA
        \and
            Instituto de Astronomıa, Universidad Nacional Autónoma de México, A.P. 70-264, 04510, Mexico, D.F., México 
        \and
            Department of Astronomy, University of Illinois at Urbana-Champaign, Urbana, IL 61801, USA 
        \and
            Department of Astronomy, University of Washington, Box 351580, Seattle, WA 98195, USA 
        \and
            Instituto de Alta Investigaci{\'{o}}n, Universidad de Tarapac{\'{a}}, Casilla 7D, Arica, Chile 
        }

   \date{Received September    , 20  ; accepted March   , 20  }
 
\abstract {
We present and discuss optical emission line properties obtained from the analysis of Sloan Digital Sky Survey (SDSS) spectra for an X-ray selected sample of 3684 galaxies ($0.002 < z < 0.55$), drawn from the eRASS1 catalog.
We modeled SDSS-V DR19 spectra using the {\sc NBursts} full spectrum fitting technique with E-MILES simple stellar populations (SSP) models and emission line templates to decompose broad and narrow emission line components for correlation with X-ray properties.
We place the galaxies on the Baldwin-Phillips-Terlevich (BPT) diagram to diagnose their dominant excitation mechanism. 
We show that the consistent use of the narrow component fluxes 
shifts most galaxies systematically and significantly upward to the active galactic nuclei (AGN) region on the BPT diagram.
On this basis, we confirm the dependence between a galaxy’s position on the BPT diagram and its ($0.2-2.3$~keV) X-ray/H$\alpha$ flux ratio.
We also verified the correlation between X-ray luminosity and emission line luminosities of the narrow [O\iii]$\lambda 5007$ and broad H$\alpha$ component; as well as the relations between the Supermassive Black Hole (SMBH) mass, the X-ray luminosity, and the velocity dispersion of the stellar component ($\sigma_{*}$) on the base on the unique sample of optical spectroscopic follow-up of X-ray sources detected by eROSITA.
These results highlight the importance of emission line decomposition in AGN classification and refine the connection between X-ray emission and optical emission line properties in galaxies.
}
  

\maketitle

\section{Introduction}
\label{sec:intro}

Since the discovery of celestial X-ray sources \citep{Giacconi1962}, X-ray astronomy has provided crucial insights into galaxy studies, particularly in the context of active galactic nuclei (AGN). While X-ray properties were historically analyzed separately, recent advancements, such as the integration of X-ray data into (Spectral Energy Distribution) SED fitting tools like  X-ray Code Investigating GALaxy Emission (X-CIGALE) \citep{Yang2020}, highlight their growing role in multiwavelength studies. Many galaxies exhibit emission lines in their optical spectra that arise predominantly from ionized gas \citep{Baldwinet1981, Veilleux1987, Kewley2002, Osterbrock2006}. When X-ray emission is observed from a galaxy, it is mainly attributed to AGN, as they are the most efficient sources of high-energy radiation, often outshining other potential contributors, such as X-ray binaries or hot gas within the galaxy. AGNs are characterized by accretion onto supermassive black holes (SMBHs) at their centres. High-energy radiation from an AGN accretion disk around a central SMBH or young hot stars can ionize and excite the gas and shocks in outflows \citep{Ferland1983, Harrison2012, Allen2008}.

In the optical spectra of such sources, multi-component emission line profiles are observed. The narrow-line region (NLR) is located at larger distances ($\sim$kpc) from the SMBH and is characterized by lower gas velocities, producing narrow emission lines (NELs). NLR profiles can be asymmetric due to outflows, which, with a sufficient contribution in flux and relative shift in radial velocity, can be considered a separate component. In contrast, the broad-line region (BLR) lies closer to the SMBH, within its gravitational influence, and features much higher gas velocities, resulting in broad emission lines (BELs) \citep{Antonucci1993, Urry1995}. Separating these components is crucial for understanding the physical and kinematic properties of AGN and their host galaxies, and also for AGN feedback processes and their effects on the surrounding interstellar medium \citep{Kauffmann2003, Fabian2012}.

One of the methods for modeling spectra is the {\sc NBursts} software package \citep{2007IAUS..241..175C,2007MNRAS.376.1033C}. It allows one to simultaneously estimate the physical (age and metallicity) and kinematic (radial velocity and velocity dispersion) properties of the stellar component using model grids \citep[E-MILES;][]{2016MNRAS.463.3409V}, the additive contribution of the AGN continuum, the kinematics of the NEL and BEL components, and also to obtain the fluxes of individual lines of each component. An important feature for dealing with large samples of objects is automatic selection and fitting of lists of emission lines falling within the studied spectral range, individually for each object. This approach allows one to obtain a fairly stable solution for the multi-parameter inverse problem. In particular, luminosity-weighted stellar population ages for host galaxies are reliably determined by {\sc NBursts} \citep{Koleva2008,2018ApJ...858...63C}.

\citet{Baldwinet1981} introduced the Baldwin-Phillips-Terlevich (BPT) diagram, using relatively closely spaced (in wavelength) forbidden/permitted optical line ratios to distinguish ionization sources (e.g., harder UV radiation from an AGN accretion disk around SMBH vs. ionization from star formation). \citet{Kauffmann2003} defined an empirical boundary above which AGN ionization tends to dominate, using the Sloan Digital Sky Survey (SDSS; \citealt{York2000}) galaxy spectra. Later, \citet{Kewleyet2006} defined a maximum starburst line, above which only active galaxies lie. 
The region in between is called "the composite" region,  populated by galaxies with ionization lines powered by both or by either processes. 

In previous work by \citep{Pulatova2024}, based on SDSS DR17 data \citep{Blanton2017, Abdurro'uf2022}, we found evidence suggesting that classifying a galaxy as an AGN or H\ii region on the BPT diagram depends on the X-ray/$\mathrm{H}\alpha$ ratio. Here, we continue our study with the goal of investigating in more detail the dependence of X-ray and optical emission from the source of ionisation (H\ii regions or accretion on the SMBH).

Now we can take advantage of a combination of recent developments in the field. First, the availability of the eROSITA eRASS1 catalog \citep{Merloni2024yCat, Merloni2024A&A} from the all-sky survey of the eROSITA (Extended ROentgen Survey with an Imaging Telescope Array \citep{Predehl2021}) instrument onboard the SRG (Spectrum-Roentgen-Gamma) satellite \citep{Sunyaev2021}. Second, the follow-up spectroscopy of the eRASS1 sources, X-ray selected galaxies to $z \sim 0.5$ with The SDSS V \citep[][]{Kollmeier2017, Kollmeier2019,Kollmeier2025a}. Finally, we use a novel data analysis technique provided by the {\sc NBursts} full-spectrum fitting code, which simultaneously fits stellar population, AGN continuum, and multi-component emission lines in optical and near-infrared spectra. 

In this study, we adopt a $\Lambda$CDM cosmology with $H_0 = 67.4$ km s$^{-1}$ Mpc$^{-1}$, $\Omega_m = 0.315$, and $\Omega_{\Lambda} = 0.685$ for distance calculations \citep{Planck2018}. All reported uncertainties correspond to 1-sigma confidence intervals. Additionally, all logarithms in this work are base 10 ($\log_{10}$).

The paper is organized as follows: in Section~\ref{sec:sample} we discuss our sample selection, in Section~\ref{sec:methods} we describe the application of {\sc NBursts} technique to SDSS-V DR19 spectra. We present our results in Section~\ref{sec:results}, and discuss them in Section~\ref{sec:discussion}, before we conclude in Section~\ref{sec:conclusions}.

\section{Data and sample selection}
\label{sec:sample}

\subsection{Data}
\label{subsec:data}

\begin{figure*}
    \centering
    \begin{subfigure}{0.47\textwidth}
        \includegraphics[width=\linewidth]{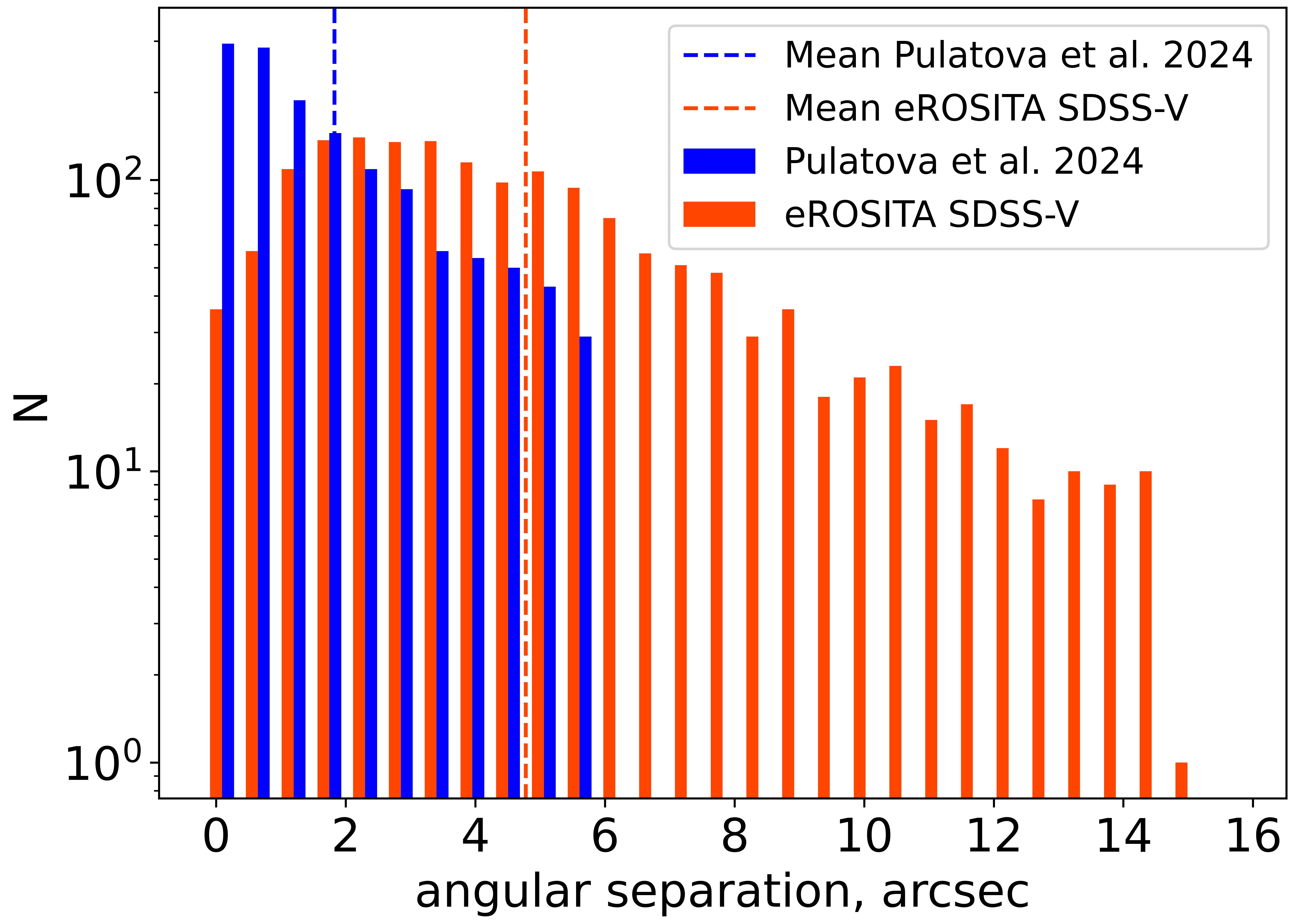}
        \caption{  }
        \label{FigAng}
    \end{subfigure}
~
    \begin{subfigure}{0.47\textwidth}
        \includegraphics[width=\linewidth]{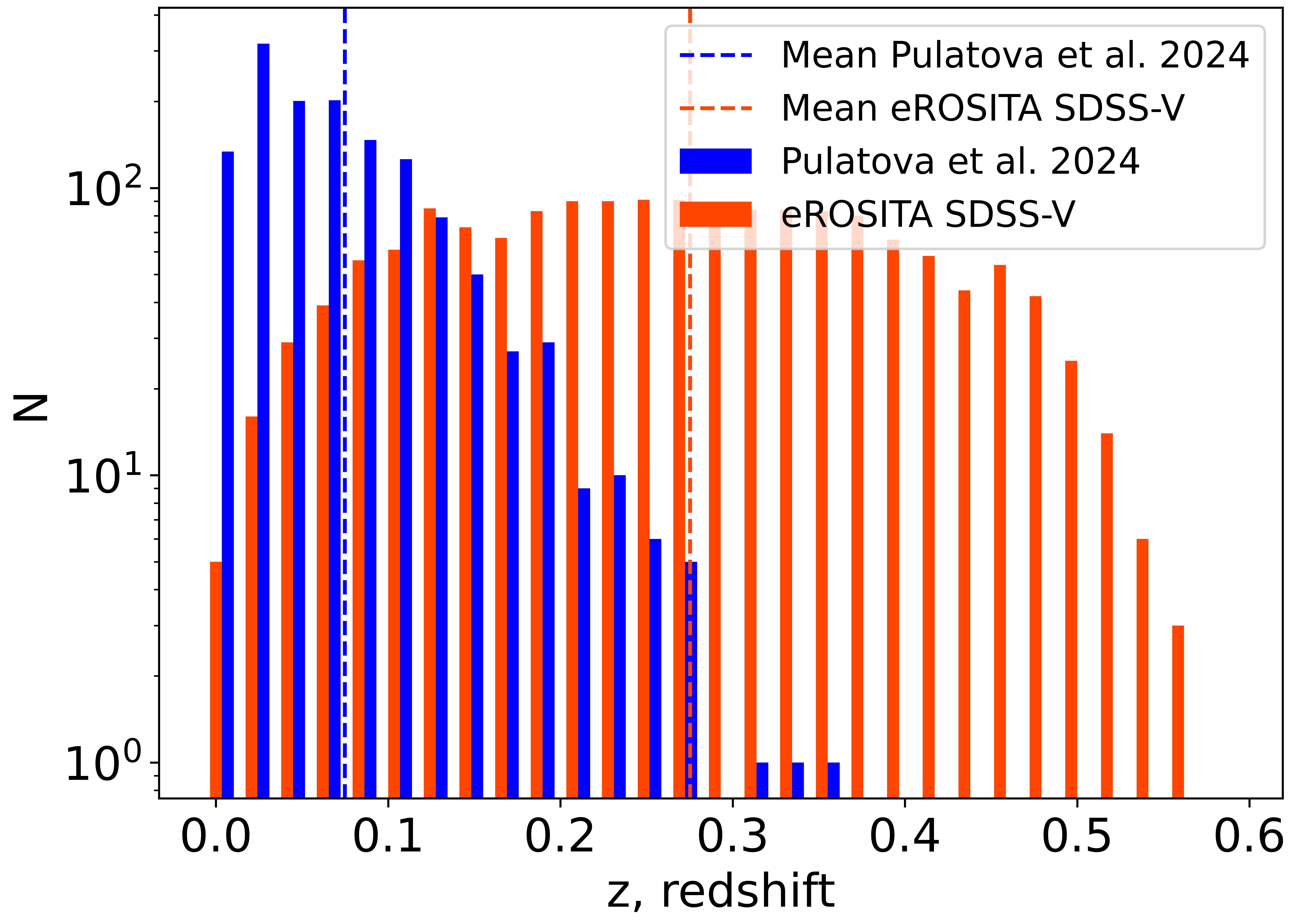}
        \caption{  }
        \label{Fig-z}
    \end{subfigure}
    
    \vskip\baselineskip
    
    \caption{
        Parameters of eROSITA and XMM-Newton X-ray selected galaxies. Panel a: angular separation between X-ray and optical sources for eROSITA and XMM-Newton X-ray selected galaxies, bin size $0.5\arcsec$. Panel b: distribution of redshift (z) for eROSITA and XMM-Newton X-ray selected galaxies, bin size $0.02$.
        }
\end{figure*}

Our study is based on the sample of eROSITA X-ray selected (eRASS1) galaxies followed up with SDSS-V (DR19). The eROSITA eRASS1 catalog provides deep, high-resolution X-ray observations of the entire sky, making it a powerful tool for studying X-ray sources such as AGN. With a sensitivity range of $10^{-14}-10^{-13}$ erg cm$^{-2}$ s$^{-1}$ in the $0.2-5$~keV band, eRASS1 can detect faint and previously unknown X-ray sources, allowing for a comprehensive analysis of the X-ray galaxy population \cite{Merloni2024A&A}. The X-ray fluxes (and derived luminosities) used in this analysis are based on the observed-frame 0.2–2.3 keV band.

SDSS-V \citep{Kollmeier2025a} is an all-sky spectroscopic survey, representing the fifth incarnation of the SDSS project \citep{Kollmeier2017, Kollmeier2019}. The program SPectroscopic IDentification of eROSITA Sources \citep[SPIDERS;][]{Dwelly2017, Comparat2019} was devised more than a decade ago, with the ultimate goal of providing SDSS optical spectra for a large number of X-ray sources detected by eROSITA~\citep{2025A&A...698A.132A}. {In this work, we use spectra from SDSS-V SDSS Data Release 19 (DR19, \cite{SDSS_Collaboration2025}), which includes all SDSS optical spectroscopic observations obtained up to mid-2022, comprising both newly acquired SDSS-V data and all legacy spectra from previous phases (SDSS-I through SDSS-IV). This allows us to work with the most comprehensive set of available SDSS spectra as of the current release.
Additionally, SDSS-V provides intermediate-resolution spectra ($R=1560-2650$) with a wavelength coverage of $3600 - 10300 \AA$, leveraging advanced data processing to maintain data quality and compatibility with previous SDSS releases.

In our paper, we use the term "correlation" to refer to visually apparent trends between parameters; no formal correlation statistics are applied, as quantifying these relationships is beyond the scope of this work.

\subsection{Sample selection}
\label{subsec:selection}

\begin{figure}
    \centering
    \includegraphics[width=\columnwidth]{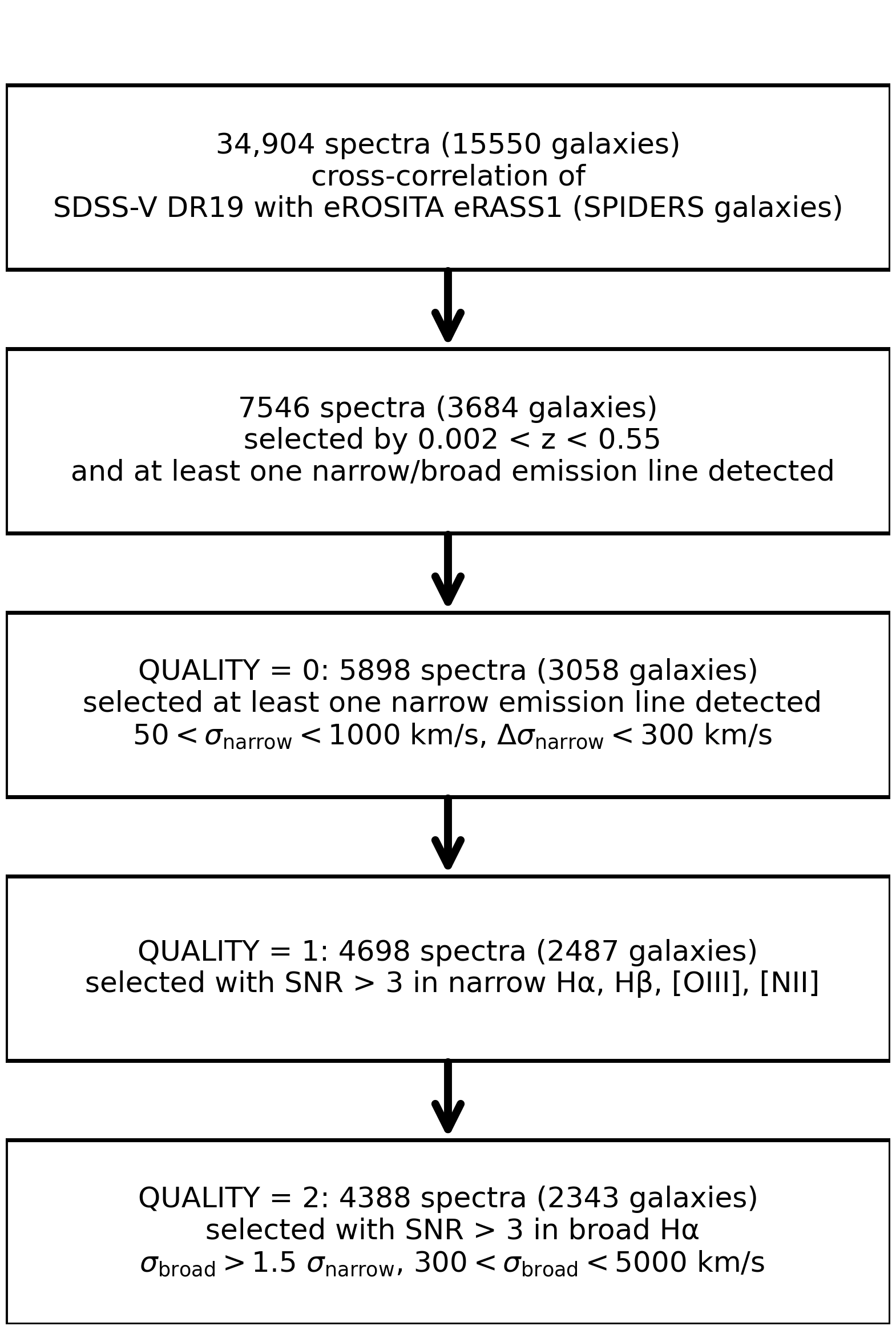}
    \caption{
        Step-by-step description of samples selection.
        }
    \label{Fig-sam-sel}
\end{figure}

The sample of X-ray–selected galaxies analyzed in this study is based on targets from the SDSS-V/SPIDERS (SPectral IDentification of eROSITA Sources) program. These sources were selected as the most likely optical counterparts to eRASS1 detections using a statistically robust procedure described in \cite{Salvato2022} and \citet{Salvato2025}. The typical $1\sigma$ position uncertainty for eROSITA sources has a mean value of $4.54\arcsec$.  The eROSITA sources were matched to the most likely optical counterparts, which then became the targets for the SDSS-V spectroscopy \citep{Kollmeier2025a}. The counterpart identification was performed using NWAY, a Bayesian cross-matching algorithm \citep{Salvato2018}, which was combined with machine-learning techniques to refine the associations. The method uses photometric and astrometric information from Legacy Survey DR8 \citep{Dey2019} and Gaia EDR3 to compute the probability that a given source is a true X-ray emitter rather than a chance alignment. This approach has been tested and shown to achieve a counterpart identification purity and completeness exceeding 93\%. In this work, we adopt the SPIDERS target list directly, rather than applying an independent matching procedure, thereby benefiting from the well-established methodology developed by the eROSITA and SDSS-V collaboration teams.
The distributions of angular separation of galaxies selected by eROSITA and XMM-Newton have mean values of $1.17\arcsec$ and $3.8\arcsec$, respectively (Fig.~\ref{FigAng}). DR19 includes the 34904 spectra of eRASS1 sources.
We then applied an additional redshift quality criterion by selecting only spectra with ZWARNING = 0, ensuring that only reliably determined redshifts were included in the sample. Additionally, we limited the redshift range to $0.002 < z < 0.55$ to exclude stars and select galaxies where the H$\alpha$ line falls within the SDSS spectral wavelength range (Fig.~\ref{Fig-z}). We also retained only spectra with detected emission lines, requiring at least one of the strong lines (H$\alpha$, H$\beta$, [O\iii]$\lambda 5007$, [O\ii]$\lambda \lambda 3727, 29$) to have a signal-to-noise ratio SNR~$> 3$. As a result, 7546 spectra of 3684 sources were selected for further analysis of emission line properties using the {\sc NBursts} full-spectrum fitting technique.

In Fig.~\ref{Fig-sam-sel} we present the stages that we followed to obtain the final sample of galaxies and the subsample of objects with the most reliable estimates of emission line parameters presented in  Section~\ref{sec:results}.

\section{Methods}
\label{sec:methods}

\subsection{{\sc NBursts} full spectrum fitting technique}
\label{subsec:NBursts}

\begin{figure*}
    \centering
    \includegraphics[width=1.\linewidth]{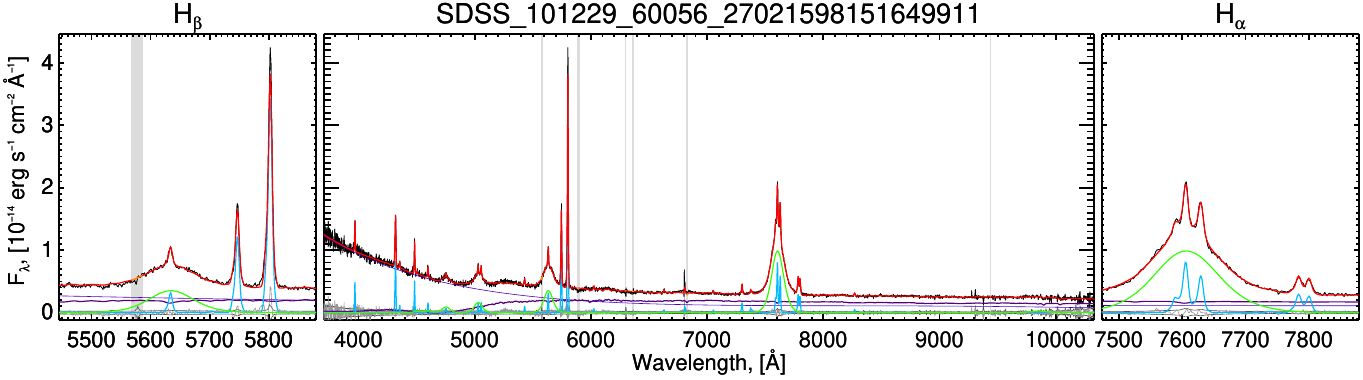}
    \includegraphics[width=1.\linewidth]{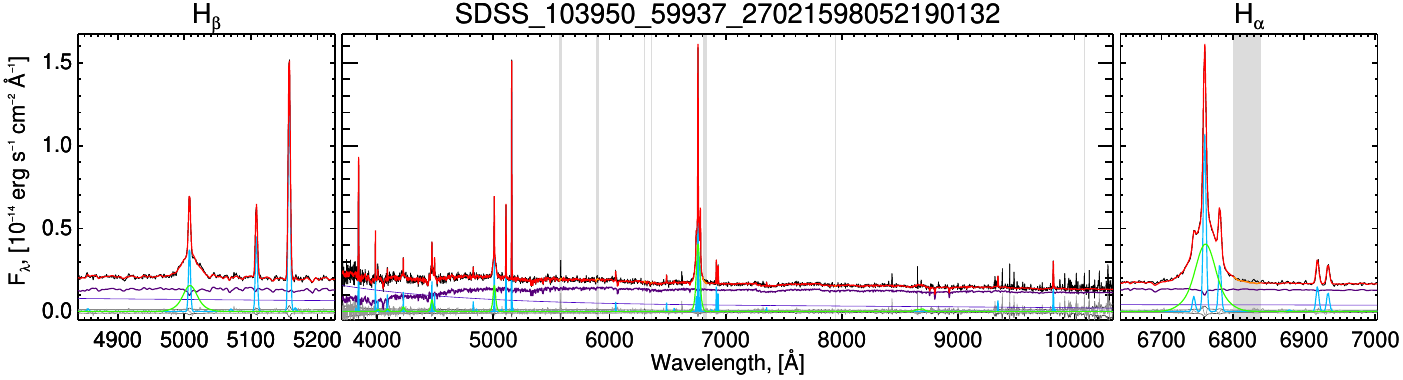}
    \includegraphics[width=1.\linewidth]{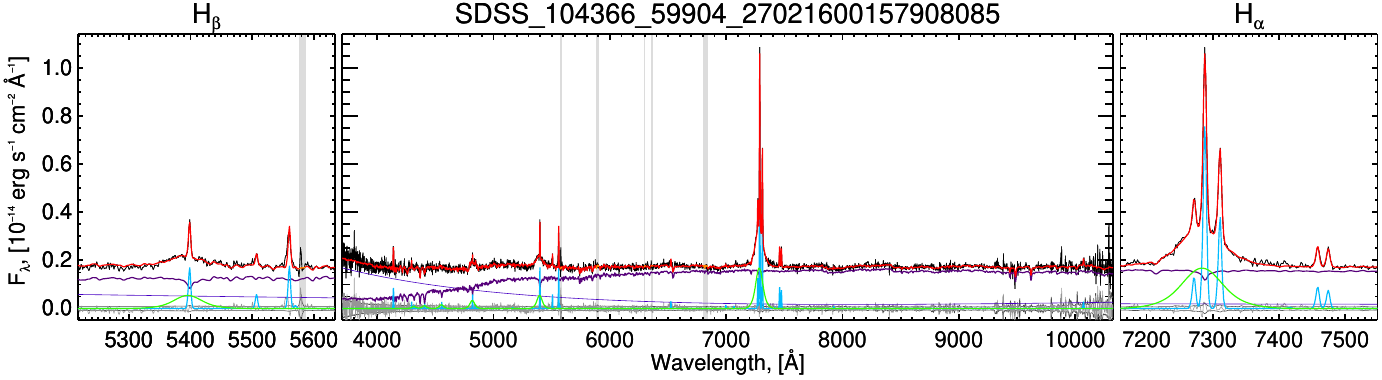}
    \includegraphics[width=1.\linewidth]{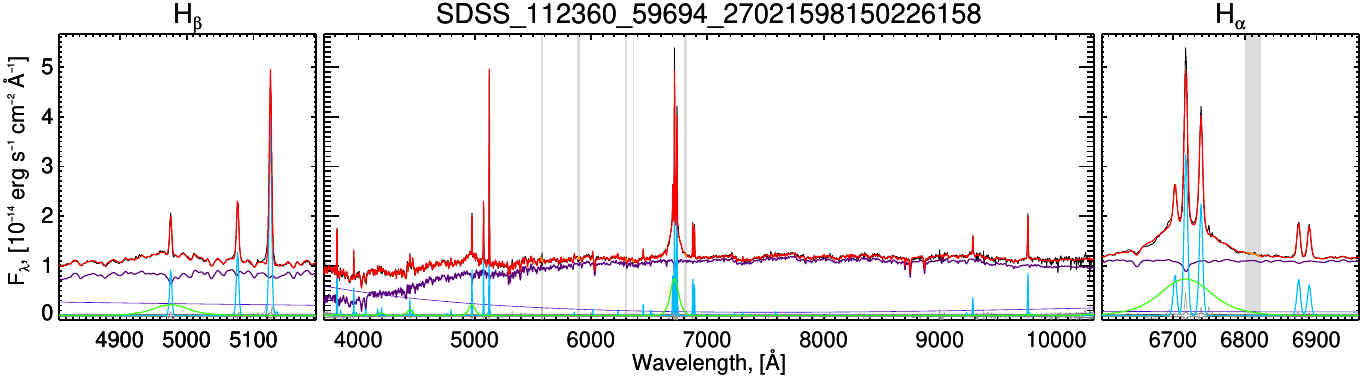}

    \caption{
        Some examples of SDSS-V spectra fitted using {\sc NBursts} full-spectrum fitting technique with E-MILES SSP models and double emission line templates (narrow + broad). Four spectra with high signal-to-noise ratio in continuum (SNR~$> 20$), which allows for a reliable identification of both narrow and broad components, were selected for demonstration. The broad-line component is clearly visible and robustly detected in multiple Balmer series lines, confirming the presence of the BLR alongside the NLR. The central panels show the spectrum in the full available wavelengths range, and the side panels show the surrounding ranges of two spectral lines H$\beta$ and H$\alpha$. The black line corresponds to spectrum fluxes, the red line --- a best-fit model, the purple line --- a stellar population model including multiplicative continuum (stellar component), the dark blue line is responsible for an additive component describing the AGN continuum, the light blue and green lines show emission lines templates for narrow and broad lines, respectively. The dark gray line shows the residuals, the light gray areas --- masked and excluded regions from the fit.
        }   
    \label{Fig-NBursts-examp}
\end{figure*}

To analyze the SDSS-V optical spectra sample from Section~\ref{subsec:selection}, we used the {\sc NBursts} full spectrum fitting technique implemented as an {\sc idl} software package \citep{2007MNRAS.376.1033C, 2007IAUS..241..175C}. We used E-MILES SSP models (\citealt{2016MNRAS.463.3409V}, $\mathrm{FWHM} = 2.5\AA$\ in the SDSS wavelength range) and automatic selection of additive templates for multi-component emission lines. The algorithm minimizes in each pixel of a spectrum the residuals between the data and the model, taking into account flux errors. The position of the $\chi^2$ minimum in the multidimensional parameter space corresponds to the best solution.
\begin{equation}
    \chi^2 = \sum_{N_{\lambda}} p_{\lambda} \frac{ \left( F_{\lambda} - F_{\lambda}^{model} \right)^2}{\Delta F_{\lambda}^2}
\end{equation}
where
$F_{\lambda}$ is the measured flux at $\lambda$ wavelength;
$F_{\lambda}^{model}$ is the model flux at the same wavelength;
$\Delta F_{\lambda}$ is the measured flux uncertainty;
$p_{\lambda}$ is the mask (0 or 1) for each pixel.

In this work, the following model is used in the minimization routine:
\begin{equation}
\begin{split}
    F^{model} = & P^{mult} \left( T^{\mathrm{SSP}} \ast \mathcal{L}^{\mathrm{SSP}} \right) +
    \sum_{N^{\mathrm{narrow}}} w^{\mathrm{narrow}} \left( T^{\mathrm{narrow}} \ast \mathcal{L}^{\mathrm{narrow}} \right) + \\
    &
    \sum_{N^{\mathrm{broad}}} w^{\mathrm{broad}} \left( T^{\mathrm{broad}} \ast \mathcal{L}^{\mathrm{broad}} \right) + 
    P^{\mathrm{AGN}}
\end{split}
\end{equation}
where
$T^{\mathrm{SSP}}$(T,[Fe/H]) is an E-MILES SSP (simple stellar population) model for given values of age (T) and metallicity ([Fe/H]) convolved according to the line-spread function (LSF) of the spectrograph (parametrized Gauss-Hermite function);
$T^{\mathrm{narrow}}$ and $T^{\mathrm{broad}}$ are the normalized emission line profiles (narrow and broad) convolved according to the LSF;
$w^{\mathrm{narrow}}$ and $w^{\mathrm{broad}}$ are the fluxes of the corresponding emission lines;
$P^{\mathrm{AGN}}$ is a 3rd order additive Legendre polynomial to take into account the contribution of the AGN continuum;
$P^{mult}$ is a low-order ($\leq 15$) multiplicative Legendre polynomial used for calibrating model fluxes and correcting small smooth variations in the continuum level resulting from raw data reduction;
$\mathcal{L}(v,\sigma,h_3,h_4)$ are parameters of the line-of-sight velocity distribution (LOSVD) for each kinematical component (SSP, narrow, broad);

Normalized physical Gauss--Hermite series for the LOSVD convolution kernels have the form:
\begin{equation}
\begin{split}
    \mathcal{L}&(v_0, \sigma, h_3, h_4) = \\
    &
    \frac{1}{\sqrt{2 \pi \sigma^2}} 
    \exp{\left[-\frac{1}{2} \left( \frac{v - v_0}{\sigma} \right)^2 \right]}
    \left( 1 + \sum_{m = 3,4} h_m H_m\left(\frac{v - v_0}{\sigma}\right) \right)
\end{split}
\end{equation}
where
$v_0$ is the radial velocity;
$\sigma$ is the velocity dispersion;
$h_3$ and $h_4$ are the Gauss-Hermite coefficients \citep{1993ApJ...407..525V} limited to $-0.15 < h < 0.15$ for normalized physical Hermite polynomials ($H_m$, \citet{2017MNRAS.466..798C}).

The emission line templates are computed individually in several steps for each spectrum. First, the lines from the list of lines (see App.~\ref{app:eml}) that fall within the spectrum wavelength range are selected based on the initial guess of the radial velocity (galaxy redshift). Then they are divided into three groups:
\begin{itemize}
    \item[i] Strong narrow emission lines that are most likely to be in the spectrum (such as H$\alpha$, [N\ii]$\lambda \lambda 6548, 84$, H$\beta$, [O\iii]$\lambda 5007$, etc.);
    \item[ii] Weak narrow emission lines, which are much less common, and their presence must be checked (each line is included in the fit only if it passes the flux criterion SNR~$> 3$);
    \item[iii] Broad component of the Balmer series emission lines (all lines are included automatically if at least one passes the criteria for flux SNR~$> 3$ and velocity dispersion $\sigma_{\mathrm{broad}} > 1.5\ \sigma_{\mathrm{narrow}}$);
\end{itemize}
A continuum fitting is performed with masks at the expected emission line positions, and then the residuals are fitted to detect lines from the groups (ii) and (iii). Finally, a fit of the original spectrum is run with stellar population models, the AGN continuum, and the entire set of detected emission lines split into narrow and broad components. To improve the stability of the solution, relative linear constraints are applied to the fluxes of individual lines based on atomic physics considerations for doublets and the hydrogen Balmer series. It is also important to note the absence of separate additive components for the Fe II lines, which are commonly observed in AGN spectra, due to the lack of widely accepted templates. However, their flux contribution is partially absorbed by additive and multiplicative continua and does not affect the kinematics of other emission lines in this spectral region, as all emission lines are fitted simultaneously across the entire available spectral range.

Thus, the {\sc NBursts} procedure yields:
\begin{itemize}
    \item Kinematic parameters (radial velocity, velocity dispersion and higher moments $h_3$, $h_4$ for the Gauss-Hermite convolution kernel) separately for each component (SSP ($h_3^{\mathrm{SSP}} = h_4^{\mathrm{SSP}} = 0$), broad, narrow components ($h_3^{\mathrm{broad}} = 0$));
    \item Properties of the stellar population (age and metallicity) based on the selected model grid (E-MILES at rest frame $3700 - 10000 \AA$\ wavelength range);
    \item Contribution of the AGN continuum (described by an additive 3rd-order polynomial);
    \item Fluxes and flux uncertainties in emission lines (separately in narrow and broad components);
\end{itemize}
Formal uncertainties in the fitted parameters are derived from the covariance matrix provided by the Levenberg–Marquardt algorithm as implemented in the MPFIT software package \citep{MPFIT}.

Several examples of spectral fits with a good SNR~$> 20$ (in continuum) are shown in Fig.~\ref{Fig-NBursts-examp}. This approach was successfully used in the past with large samples of spectra, for example, in the \href{http://rcsed.sai.msu.ru}{RCSED}\footnote{\url{http://rcsed.sai.msu.ru}} \citep{Chilingarian2017} project for the SDSS DR7 spectra, and is used in the form described above in the next generation of the \href{https://rcsed2.voxastro.org/}{RCSEDv2}\footnote{\url{https://rcsed2.voxastro.org/}} project with spectral collections from a dozen spectral sky surveys. To verify the reliability of the results obtained in Section~\ref{sec:results}, we also compared them with the RCSED and RCSEDv2 databases. Best-fitting results from previous SDSS data releases are publicly available through the interactive web-service RCSEDv2 (an updated version of the value-added Reference Catalog of Spectral Energy Distributions of galaxies); further details can be found about the catalog of galaxy properties in \citet{2024ASPC..535..179C}, the spectrum visualization service in \citet{2024ASPC..535..243K}, the processing of spectra in \citet{2024ASPC..535..175G}, the hybrid minimization algorithm for spectrum fitting in \citet{2024ASPC..535..371R}, and the workflow for automatic massive parallel analysis of spectra in \citet{2024IAUGA..32P2683R}.

\subsection{Black hole mass estimates}
\label{subsec:mbh}

By decomposing the permitted emission lines into broad and narrow components, we can estimate a single-epoch black hole mass ($M_{\mathrm{BH}}$) using a virial approximation applied to a broad H$\alpha$ emission line \citep{Reines2013}. This method assumes that the gas in the BLR is virialized and its motions are driven by the gravitational influence of the black hole. The virial mass is calculated using the velocity dispersion of the gas, which is derived from the width, FWHM, of the broad H$\alpha$ line, and the radius of the BLR, which is inferred from empirical scaling relations based on the luminosity of the broad H$\alpha$ emission, using the following equation:
\begin{equation}
\begin{split}
    \log & \left( \frac{M_{\mathrm{BH}}}{M_{\odot}} \right) =  \\
    & 6.57 + 0.47 \log \left( \frac{L_{\mathrm{H}\alpha}^{\mathrm{BLR}}}{10^{42} [{\rm erg}\, {\rm s}^{-1}]} \right) + 2.06 \log \left( \frac{\mathrm{FWHM}_{\mathrm{H}\alpha}^{\mathrm{BLR}}}{10^3 [{\rm km}\, {\rm s}^{-1}]} \right),
\end{split}
\end{equation}
where the coefficients derived by \citet{Reines2013} are from the $R_{\mathrm{H}\beta}-L_{5100}$ relation \citep{2013ApJ...767..149B} and the $\mathrm{FWHM}_{\mathrm{H}\beta}-\mathrm{FWHM}_{\mathrm{H}\alpha}$, $L_{\mathrm{H}\alpha}-L_{5100}$ relations from \citet{2005ApJ...630..122G}.

For the subsample of spectra with reliable broad line component detections (SNR>3 in broad $H\alpha$, $\sigma_{broad}$ > 1.5 $\sigma_{narrow}$ and $\sigma_{broad}$ is in range from 300 to 5000 km/s, QUALITY=2, see Sec.~\ref{subsec:table}, and Fig.~\ref{Fig-sam-sel}), the masses of black holes were estimated. We consider the measurement uncertainties of $\mathrm{FWHM}_{\mathrm{H}\alpha}^{\mathrm{BLR}}$ and broad $\mathrm{H}\alpha$ flux returned by the {\sc NBursts} code, which we propagate to estimate $M_{\mathrm{BH}}$ uncertainties. 

\subsection{Fitting of empirical power-law relations}
\label{subsec:linfit}

To fit empirical relations between parameters [$L_{X}$ vs $L_{OIII}$] as power laws, we used the nested sampling algorithm \citep{10.1063/1.1835238,2023StSur..17..169B} implemented in the Python package {\sc UltraNest} \citep{2021JOSS....6.3001B}.
The likelihood is the line (in log space) with slope, offset, and intrinsic scatter; priors are normal distributions.
The posteriors for slope, offset, and intrinsic scatter for the $L_{X}(L_{OIII})$ relation, are available on Zenodo~\citep{zenodo_lxlo}\footnote{\url{https://zenodo.org/records/16520014}}. 
Graphs with linear fits through the paper represent ordinary least-squares fits.

\section{Results}
\label{sec:results}

\subsection{Main results and sub-samples}
\label{subsec:table}

The final fitting results for 7546 spectra of 3684 sources from the SDSS-V DR19 eROSITA eRASS1 X-ray selected galaxies sample in the $0.002 < z < 0.55$ range are presented in a table available at the CDS (see Appendix~\ref{app:tab} for column description). The table contains lists of the main emission lines in the rest frame wavelength range $3700 - 7000 \AA$ from [O\ii]$\lambda 3727$ to [S\ii]$\lambda 6731$ --- 28 lines for the narrow component and 6 lines of the Balmer series (from H8 to H$\alpha$) for the broad component. The absence of any line identifier in the specified wavelength range in the list of lines does not necessarily indicate its absence in each individual spectrum. A total of 140 lines fell within the fitting ranges, from which we selected 34 lines for the final compiled table, which occurred in spectra most frequently. X-ray fluxes at different energies from eRASS1 based on our cross-match were included in the table. Finally, we included quality flags for the fitting results based on the set of parameters obtained; see the QUALITY field, which can take four values according to the following filters:

\begin{itemize}
\item[0:] A subsample with reliable detection of a narrow component comprising at least 3 forbidden unblended emission lines with SNR~$> 3$ (e.g. [O\iii], [O\ii], etc.) and fulfilling the kinematic constraints: $50 < \sigma_{\mathrm{narrow}} < 1000$~km/s and $\Delta \sigma_{\mathrm{narrow}} < 300$~km/s;

\item[1:] A subsample of (0) for BPT diagram classification with additional conditions of $\mathrm{SNR} > 3$ in each of the following lines: H$\alpha$, H$\beta$, [N\ii]$\lambda \lambda 6548, 84$, [O\iii]$\lambda5007$;

\item[2:] A subsample of (1) with reliable detection of a broad component: $\mathrm{SNR}_{\mathrm{H}\alpha}^{\mathrm{broad}} > 3$ and $\sigma_{\mathrm{broad}} > 1.5\ \sigma_{\mathrm{narrow}}$, $300 < \sigma_{\mathrm{broad}} < 5000$~km/s, $\Delta \sigma_{\mathrm{broad}} < 300$~km/s;

\item[-1:] All other spectra that do not fall into any of the above categories are marked as unreliable;
\end{itemize}

In the final subsamples:
(i) with QUALITY = 0 we have 5898 spectra of 3058 sources with reliable measurement of narrow-line properties [2844 of these have broad lines detected according to the criteria stated in (2)]; 
(ii) with QUALITY = 1 we have 4698 spectra of 2487 sources suitable for the BPT classification; 
(iii) with QUALITY = 2 we have 4388 spectra of 2343 sources with the reliable detection of a broad component and suitable for black hole mass estimates (method described in Sec.~\ref{subsec:mbh});
(iv) The remaining 1648 spectra of 1130 sources have QUALITY = -1. We remark that this final subset is not completely hopeless; they may be useful for other tasks related to the study of individual emission line properties, but this requires an appropriate selection of additional criteria for emission-line measurements.

The sequence of obtaining subsamples is depicted in Fig.~\ref{Fig-sam-sel}. For further diagnostics and verification of dependencies, we focus herein on a subsample of sources with the QUALITY = 2 flag as the most reliable candidates with sufficient signal (SNR~$> 3$) in the main broad and narrow emission lines. 
For users interested in analyzing sources with a broader range of spectral quality, the selection criterion  QUALITY > 0 can be used to include all objects with acceptable (non-zero) spectral classifications.

\subsection{Emission lines comparison with previous surveys}
\label{subsec:emis}

\begin{figure*}
    \centering
    \begin{subfigure}{0.47\textwidth}
        \includegraphics[width=\linewidth]{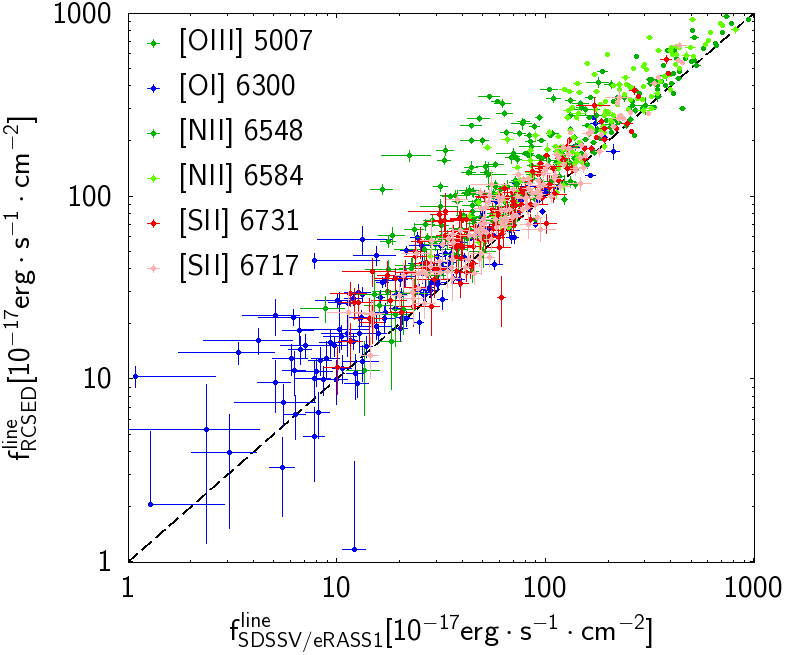}
        \caption{    }
        \label{Fig-line-decom-a}
    \end{subfigure}
~ 
    \begin{subfigure}{0.47\textwidth}
        \includegraphics[width=\linewidth]{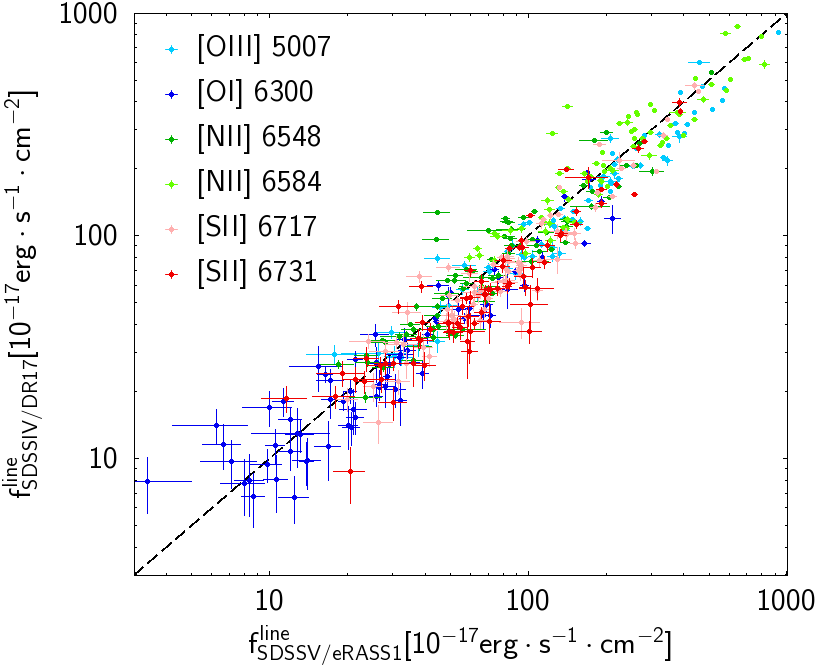}
        \caption{    }
        \label{Fig-line-decom-broad-a}
    \end{subfigure}

    \vskip\baselineskip

    \caption{
        Comparison of fluxes in forbidden optical emission lines obtained with {\sc NBursts} technique for X-ray selected DR19 galaxies (x-axis, this paper) with previously published results (y-axis). The left panel compares 101 galaxies in common with the RCSED catalog \citep{Chilingarian2017}. The right panel --- 75 galaxies in common with RCSED SDSS DR17 data release \citep{Blanton2017}. The dashed black line represents 1:1 relation. The fluxes obtained in this work showed a good agreement with SDSS-IV DR17 data and a correlation with RCSED.
        }
    \label{Fig-line-decom}
\end{figure*}

\begin{figure}
    \centering
    \includegraphics[width=\columnwidth]{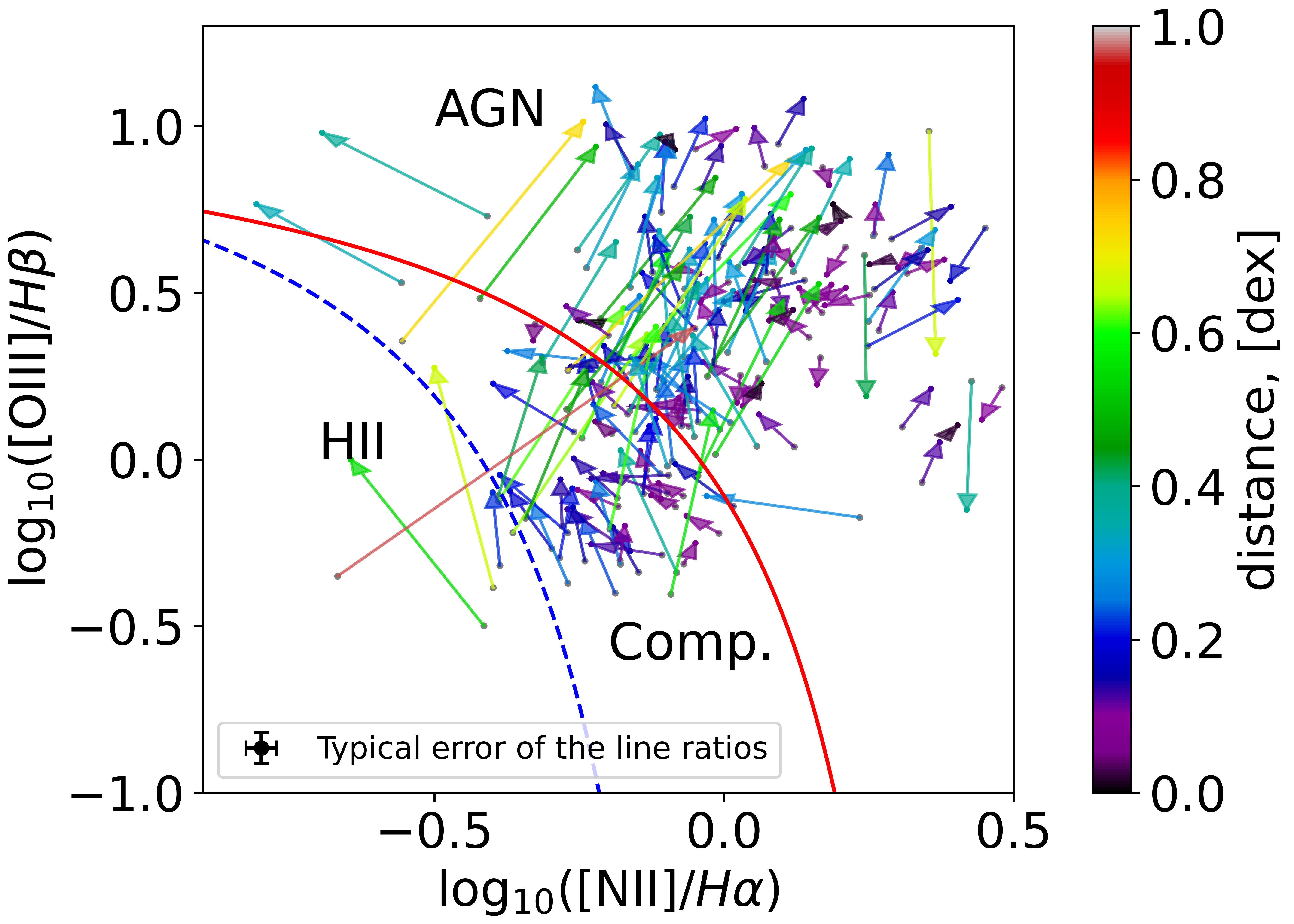}
    \caption{
        Change of position on the BPT diagram for 101 galaxies of the matched sample from this work with the RCSED catalog. At the base of the arrow is the position determined by the emission line fluxes from the RCSED without decomposition, and at the top of the arrow is the position determined by the narrow component as a result of decomposition into narrow + broad components. Color code denotes the ``distance'' between the two measures in the diagram.
        }
    \label{Fig-BPT1-up}
\end{figure}

To validate our results for measurements of optical emission line fluxes, we compared them with previously published fluxes from RCSED \citep{Chilingarian2017} and SDSS-IV DR17 \citep{Blanton2017}. The RCSED project is based on the analysis of  SDSS DR7 spectra using an earlier version of the {\sc NBursts} code without emission line decomposition. The SDSS-IV DR17 catalog also provides emission line fluxes without decomposition into narrow + broad components. To perform this test, we cross-matched these catalogs with the QUALITY = 2 flag X-ray sample using {\sc TOPCAT} with an angular separation of $1 \arcsec$ and requiring SNR~$> 3$ in the emission lines. We found 101 and 75 sources in the two groups, respectively. 
It showed a good agreement
 with SDSS-IV DR17 data and  correlation with RCSED (see Fig.~\ref{Fig-line-decom} Sec.~\ref{subsec:emis})

As a further comparison, we examine the BPT diagram locations in Fig.~\ref{Fig-BPT1-up} for RCSED (base of the arrow), where no decomposition is made, and our line fluxes (top of the arrow), where narrow + broad components are decomposed.
The decomposition shifts the sources on the BPT diagram upwards along the $\log$[O\iii]$/\textrm{H}\beta$ axis in 82\% of cases, with only 18\% shifted downwards with a median shift of +0.17 dex and a 16–84\% percentile
range of [+0.07, +0.38] dex. The analysis highlights the importance of the decomposition in accurately identifying whether the ionization mechanism is dominated by AGN or star formation and in preventing misclassifications that arise when broad components from the AGN are included in the total flux. We notice that the broad component detection can be caused not only by the presence of a type-I AGN, but also by high-velocity galactic winds, associated shocks \citep{2009ApJ...701..955S} or internal kinematics, like bars \citep{Pulatova2015}.

\subsection{BPT diagrams}
\label{subsec:bpt}

\begin{figure*}
    \centering
    \begin{subfigure}{0.46\textwidth}
        \includegraphics[width=\linewidth]{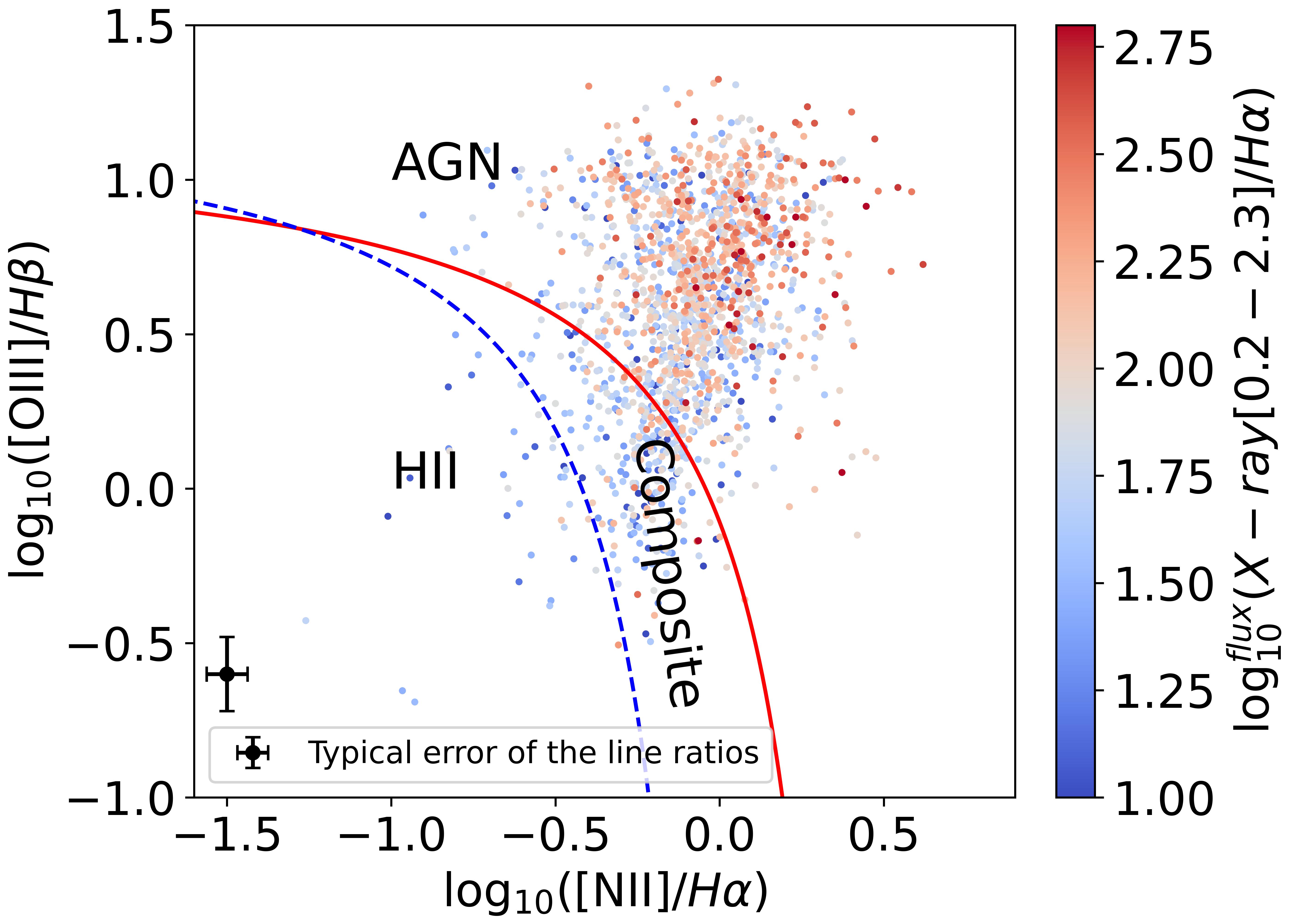}
        \caption{  }
        \label{Fig-bpt-sdss-v-X-A}
    \end{subfigure}
~
    \begin{subfigure}{0.47\textwidth}
        \includegraphics[width=\linewidth]{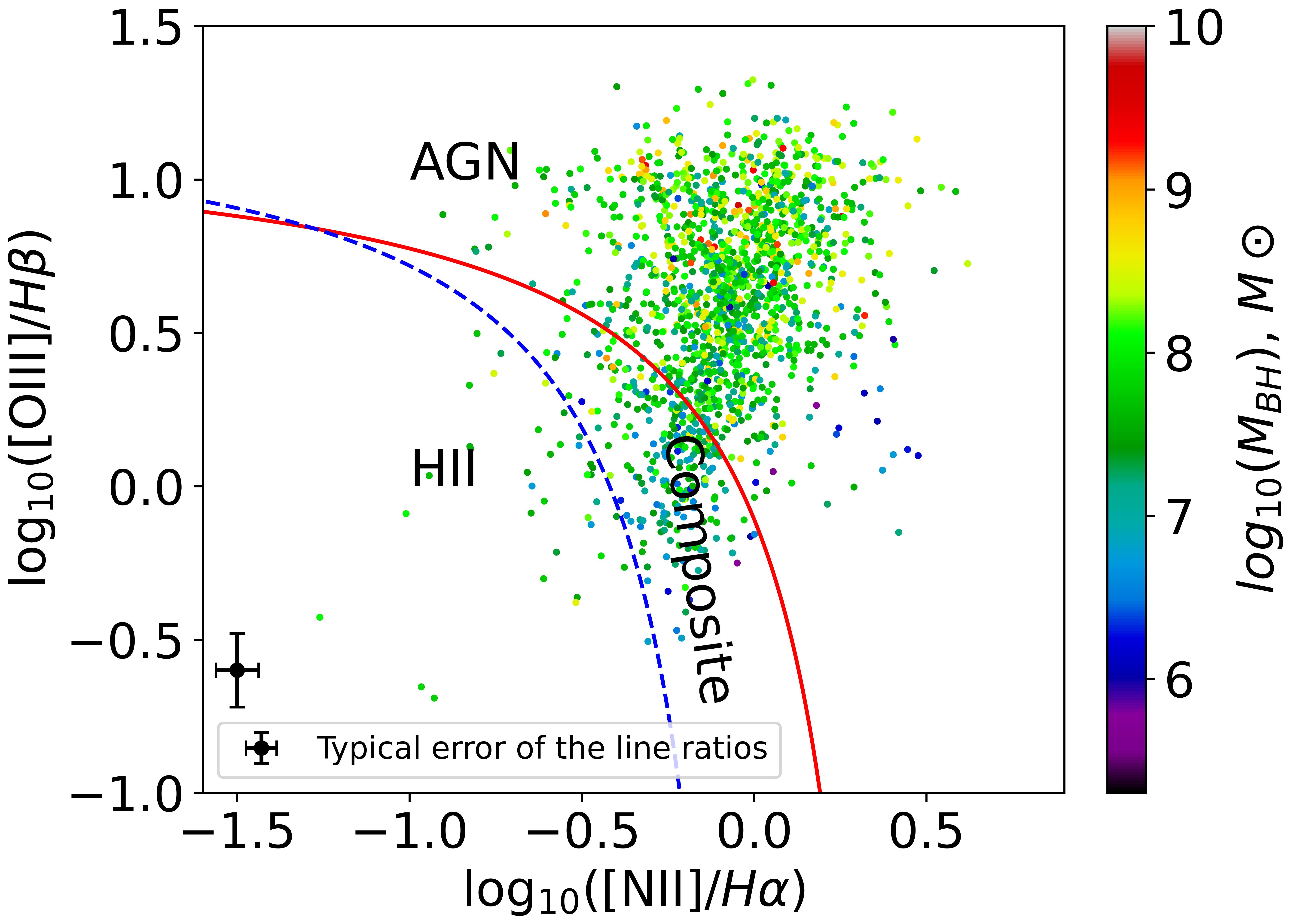}
        \caption{  }
        \label{Fig-bpt-sdss-v-m-BH}
    \end{subfigure}

    \vskip\baselineskip

    \begin{subfigure}{0.47\textwidth}
        \includegraphics[width=\linewidth]{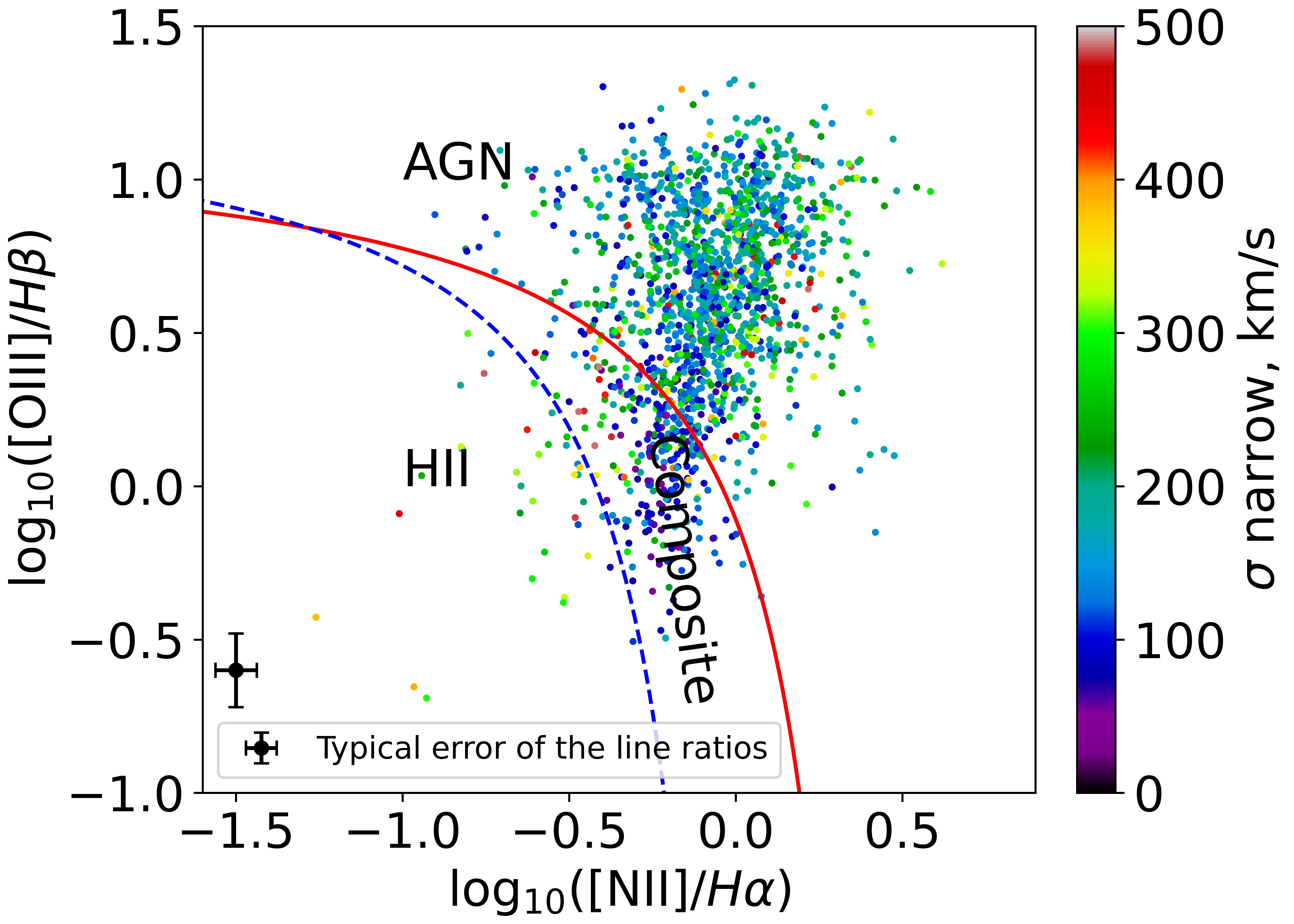}
        \caption{  }
        \label{Fig-bpt-sigma}
    \end{subfigure}
~
    \begin{subfigure}{0.47\textwidth}
        \includegraphics[width=\linewidth]{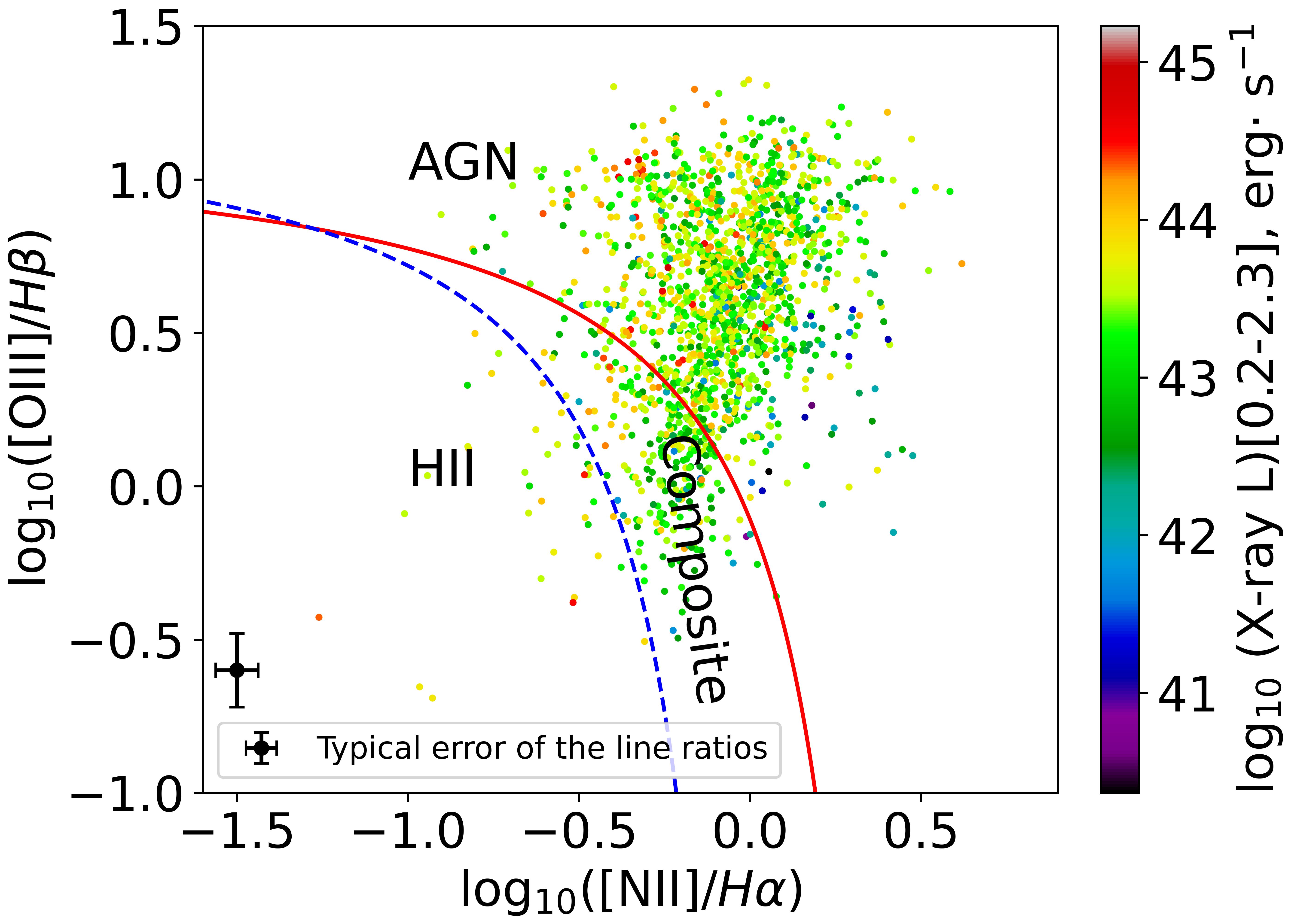}
        \caption{  }
        \label{Fig-BPT1-X-lum}
    \end{subfigure}

    \caption{
        A BPT diagram for X-ray-selected galaxies. Color code: Panel a: ratio of X-ray [$0.2 - 2.4$~keV] to H$\alpha$; Panel b: BH mass; Panel c: $\sigma_{\rm narrow}$,~km~s$^{-1}$; Panel d: X-ray luminosity in $0.2 - 2.3$~keV. 
        }
    \label{Fig-bpt-RCSED}
\end{figure*}

Having demonstrated the advantages of line decomposition to isolate AGN dominant components, we now investigate potential trends across the BPT of X-ray selected galaxies among several parameters of interest (color-coded symbols) as follows:

\begin{itemize}
\item Figure~\ref{Fig-bpt-sdss-v-X-A} is color-coded by X-ray-to-H$\alpha$ flux ratio, allowing us to confirm and strengthen the correlation found in \citet{Pulatova2024}, Fig.~6a.
Also, we see the difference in the X-ray to H$\alpha$ flux ratio at the BPT diagram for AGN in \citet{Pulatova2024} sample ($F_{\textrm{X-ray}}/F_{\textrm{H}\alpha} > 1$) and the current eRASS1 sample ($F_{\textrm{X-ray}}/F_{\textrm{H}\alpha} > 2$). The higher $F_{\textrm{X-ray}}/F_{\textrm{H}\alpha}$ ratio for AGN in the current work can be explained by the fact that for the eRASS1 sample, the line fluxes arise only from the narrow emission line component, while \citet{Pulatova2024} used total H$\alpha$ fluxes without decomposition (Sec.~\ref{sec:discussion}).

\item Figure~\ref{Fig-bpt-sdss-v-m-BH} is color-coded by single-epoch SMBH mass estimates, 
illustrating that the SMBH mass correlates with the position of a galaxy on the BPT diagram. Galaxies with SMBH masses $> 10^8 M_{\odot}$ are located mainly in the upper part of the loci of Seyfert and Low-Ionization Nuclear Emission-line Regions (LINER) on the BPT diagram. Meanwhile, galaxies with lower-mass SMBHs are more often located at the bottom of the Seyfert and LINERs loci.

\item Figure~\ref{Fig-bpt-sigma} is color-coded by the velocity dispersion ($\sigma_{\rm narrow}$) of the narrow emission line component,
to reflect the dynamics of the NLR gas and its role in influencing the position of a galaxy on the BPT diagram. Here we find that galaxies with higher $\sigma_{\rm narrow}$ are located at the top of the diagram.

\item Figure~\ref{Fig-BPT1-X-lum} is color-coded by the X-ray luminosity ($L_{0.2-2.3 keV}$).
Against expectations, we do not see a strong trend from the HII to AGN regions.
Nevertheless, galaxies with X-ray luminosities below $10^{41}$ erg~s$^{-1}$, are predominantly located in the bottom part of LINERs, suggesting that low X-ray luminosity AGN are more likely associated with weaker ionization mechanisms. This highlights the complexity of the AGN-host galaxy interactions and the need for multiwavelength approaches to better understand these systems.
\end{itemize}

A similar optical study using SDSS-V AGN data, including the WHAN diagram, was presented by \citet{Cortes-Suarez2022}. In that work, the authors emphasized the challenges of measuring narrow emission lines in spectra dominated by intense broad-line emission, where the narrow components may be partially blended or suppressed.

\subsection{X-ray [0.2-2.3] keV and optical lines luminosities}
\label{subsec:xray}

\begin{figure*}
    \centering
    \begin{subfigure}{0.49\textwidth}
        \includegraphics[width=\linewidth]{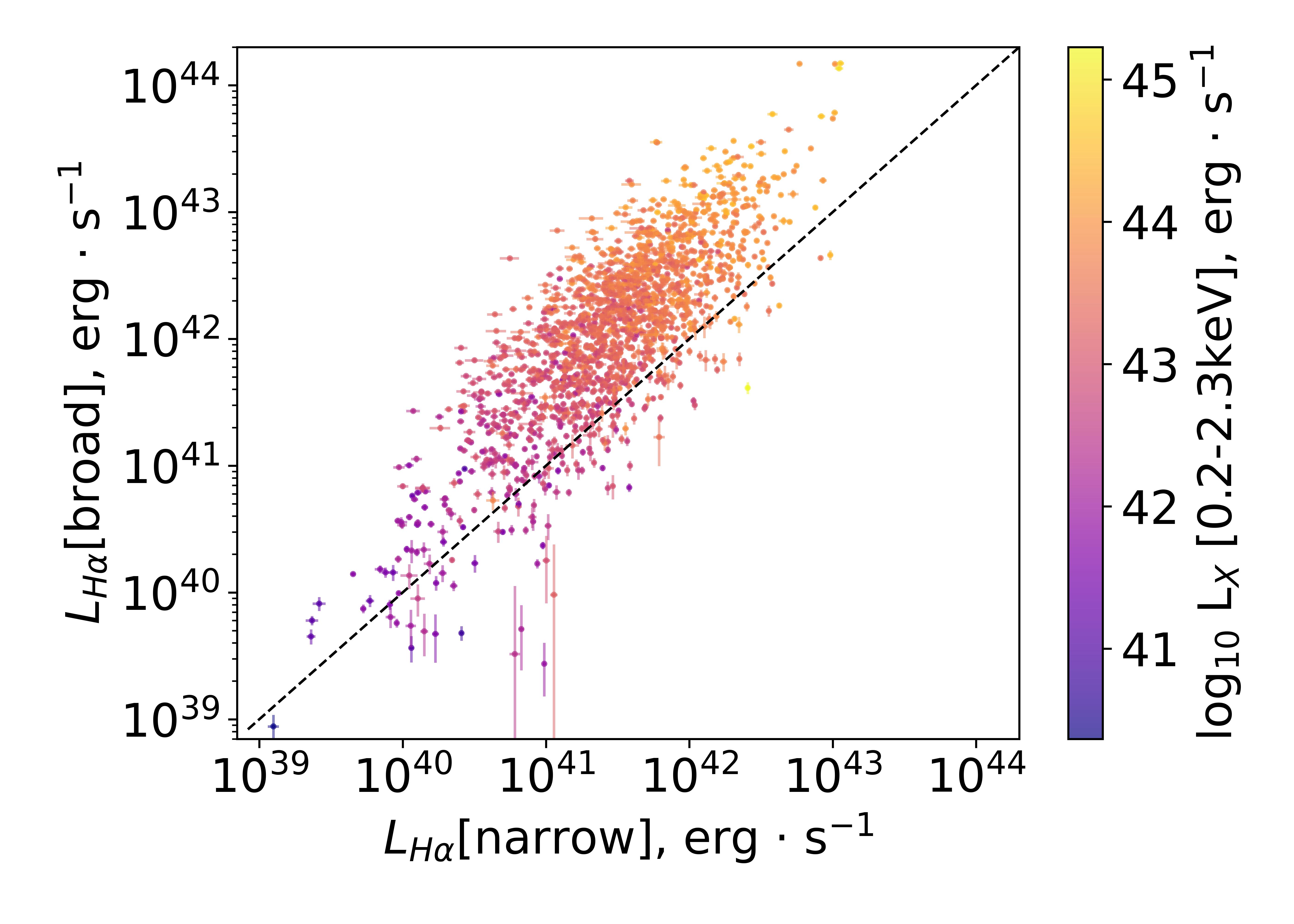}
        \caption{  }
        \label{Fig-L_Halpha_b_n}
    \end{subfigure}
~
    \begin{subfigure}{0.49\textwidth}
        \includegraphics[width=\linewidth]{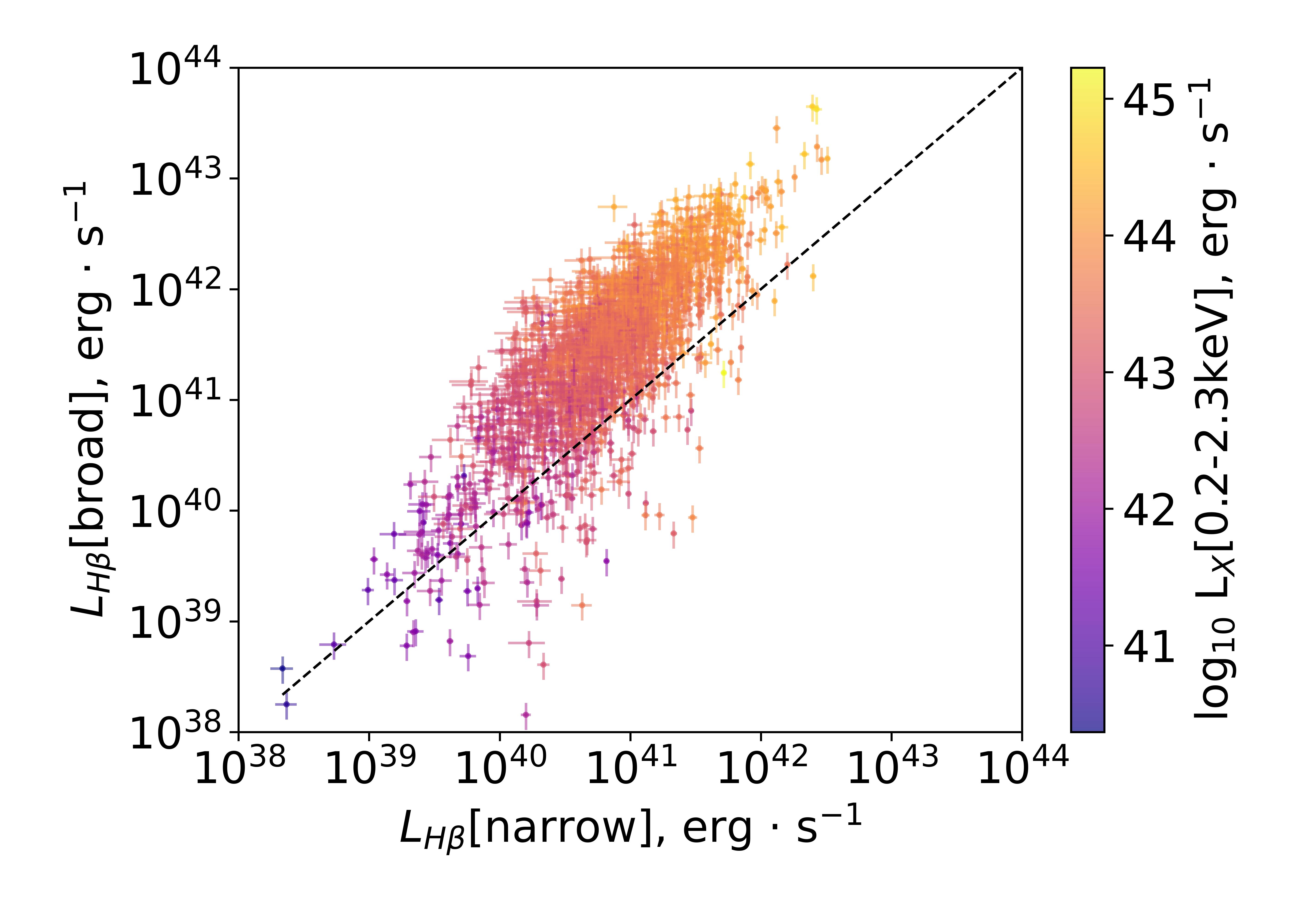}
        \caption{  }
        \label{Fig-L_Hbeta_b_n}
    \end{subfigure}

    \vskip\baselineskip

    \caption{
        Optical luminosity in the broad and narrow components. Color code: eROSITA X-ray luminosity in $0.2 - 2.3$~keV. The black dashed line represents 1:1 relation. It demonstrates a strong connection between the BLR, NLR, and the X-ray source. Panel a: H$\alpha$; Panel b: H$\beta$.
        }
    \label{Fig-Lum-nar-broad_X}
\end{figure*}

\begin{figure*}
    \centering
    \begin{subfigure}{0.49\textwidth}
        \includegraphics[width=\linewidth]{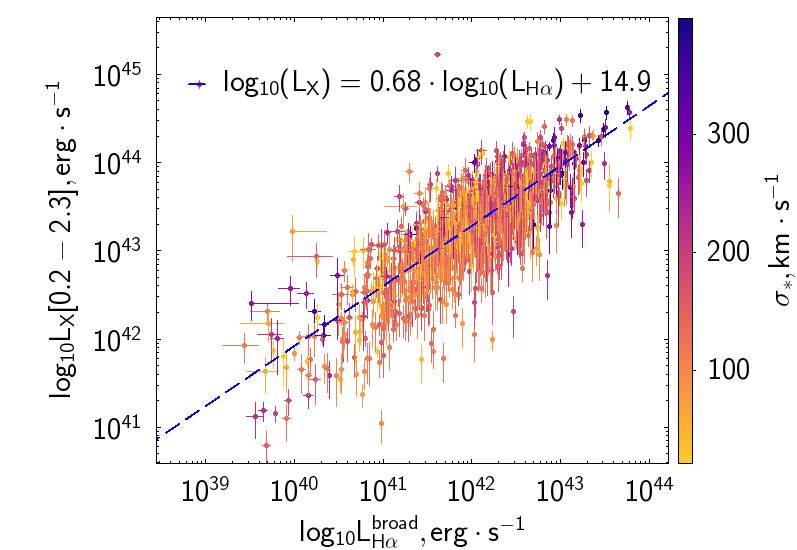}
        \caption{  }
        \label{Fig-L_Halpha_L_X}
    \end{subfigure}
~
    \begin{subfigure}{0.475\textwidth}
        \includegraphics[width=\linewidth]{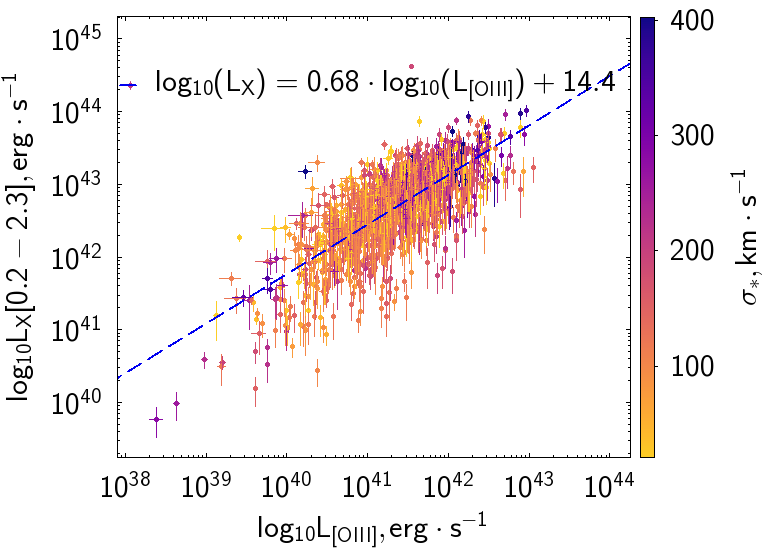}
        \caption{  }
        \label{Fig-L_OIII_L_X}
    \end{subfigure}

    \vskip\baselineskip

    \caption{
        Correlations between the X-ray [$0.2-2.3$~keV], broad H$\alpha$ (a) and narrow [O\iii] (b) luminosities. The color coding represents the stellar component's velocity dispersion, linking the kinematics to AGN activity. The dashed lines and equations near the top of the panel denote the best-fit correlations. The uniform scaling between X-ray and line luminosities reinforces their common AGN origin.
        }
    \label{Fig-lum-[OIII]-Halpha}
\end{figure*}

\begin{figure}
    \centering
    \includegraphics[width=\columnwidth]{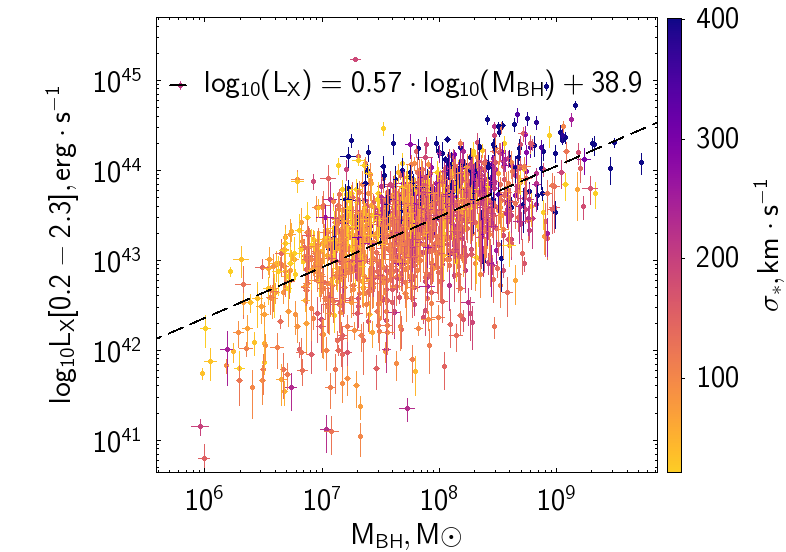}
    \caption{
        The X-ray luminosity --- SMBH mass diagram. Color code: velocity dispersion of the stellar component $\sigma_{*}$.
        }
    \label{Fig_M_BH_LX}
\end{figure}

In Fig.~\ref{Fig-Lum-nar-broad_X} we compare the broad and narrow line luminosities for H$\alpha$ (a) and H$\beta$ (b), respectively. The eROSITA X-ray luminosity ($0.2-2.3$~keV range) is color-coded. We do not detect a significant difference between H$\alpha$ and H$\beta$. The X-ray luminosity correlates with both the broad and narrow components of these Balmer lines, and the trends are consistent. One notable difference is the relative flux contribution, which we explain below: for the majority of galaxies in our sample, the flux in the broad component is higher than that in the corresponding narrow component. This trend is clearly visible in Fig.~\ref{Fig-Lum-nar-broad_X}, where most data points lie above the one-to-one line (x=y). Only a small fraction of sources show the opposite behavior, with the narrow component dominating over the broad. We also note that at lower X-ray fluxes, a larger fraction of the Balmer narrow-line emission may originate from star formation in the host galaxy rather than from AGN photoionization. This is especially relevant in systems where the AGN contribution to the total ionizing budget is modest or obscured. In such cases, the narrow components may reflect H\ii region emission rather than excitation from the AGN NLR, contributing to additional scatter in the correlation between narrow-line luminosity and X-ray emission.
This trend is consistent with previous findings \citep{Kauffmann2003, Ho2008}. The correlations in Fig.~\ref{Fig-Lum-nar-broad_X} are physically motivated. The broad components originate in the BLR, which lies very close to the central SMBH and is directly exposed to the ionizing continuum. In contrast, the narrow components are produced in the more extended NLR, where local gas density, obscuration, and ionization geometry can vary significantly. The fact that we observe consistent correlations for both components suggests that, despite differences in spatial scale and physical conditions, the optical emission lines still reflect the ¨strength¨ of the central ionizing source.

To further explore the physical connection between X-ray emission and ionized gas properties, Fig.~\ref{Fig-lum-[OIII]-Halpha} shows the correlation between X-ray luminosity in the 0.2–2.3~keV band and the luminosities of two prominent emission lines: H$\alpha$ (broad component; panel a) and [O\iii] $\lambda$5007 (panel b). The stellar velocity dispersion ($\sigma_{*}$) is color-coded in both panels. These emission-line luminosities, while often used as observational tracers of AGN activity, are influenced by the ionizing UV continuum and the structure of the emitting regions. Similar correlations have been discussed in previous studies \citep{Panessa2006}, and our results extend this picture using a uniformly selected eRASS1 sample.
These panels reveal a strong correlation between X-ray luminosity and the broad component of H$\alpha$ and [O\iii], reinforcing the close connection of both emissions to the central engine of AGN. 
The best linear correlations are expressed as 
\begin{equation}
    \log L_X = (0.68 \pm 0.06) \cdot \log \textrm{H}\alpha_{\textrm{broad}} + (14.26 \pm 0.27)
\end{equation}
and
\begin{equation}
    \log L_X = (0.68 \pm 0.02) \cdot \log [OIII] + (14.41 \pm 0.59)
\end{equation} 
The correlations presented in Equations 5 and 6 are consistent with physical models of AGN structure and emission mechanisms. Specifically, the correlation between X-ray and [O\iii] luminosities reflects the ionization of NLR gas by high-energy photons from the accretion disk and corona \citep{Heckman2014, Panessa2006, Ueda2015}. The similar correlation observed between X-ray and broad H$\alpha$ luminosity arises from photoionization of the BLR, located in the immediate vicinity of the supermassive black hole and directly illuminated by the ionizing continuum \citep{Panessa2006, Imanishi1999}. 
These approximations yield identical coefficients $0.68\pm 0.06$ (slope) and $\approx 15\pm 0.59$ (intercept) and indicate that X-ray luminosity scales uniformly with the optical luminosities of the H$\alpha$ broad component and [O\iii] emission lines. The identical coefficients suggest that the physical processes governing the generation of X-ray and optical, BEL, and NEL emissions are closely linked, regardless of the emission line used. 
This similar scaling observed for both H$\alpha$ (broad component) and [O\iii] (narrow component) suggests a physical connection between the central AGN activity, which powers the X-ray emission, and the ionized gas in both the BLR and NLR.

Fig.~\ref{Fig_M_BH_LX} examines the relationship between X-ray luminosity  ($0.2 - 2.3$~keV) and SMBH mass for X-ray galaxies, with the stellar velocity dispersion color-coded. 
As expected, we observe a general trend of increasing X-ray luminosity with SMBH mass, largely reflecting the underlying $M$–$\sigma$ relation and its intrinsic scatter \citep[e.g.,][]{Tremaine2002, Kormendy2013}. It demonstrates how higher X-ray luminosity tends to be associated with more massive SMBHs, aligning with the understanding that more massive SMBHs, on average, have higher absolute accretion rates. 
However, the distribution of Eddington rates is folded into this correlation and, additional selection effects—particularly with distance—plan be considered in our future studies. For example, small central black holes in the intermediate-mass regime \citep{Reines2013,2018ApJ...863....1C} can sometimes have X-ray luminosities in the $10^{42}-10^{43}$~erg~s$^{-1}$ range. The best-fitting linear regression has the form: 
\begin{equation}
    \log L_X = (0.57\pm 0.02)\cdot \log M_{\textrm{BH}} + (38.90\pm 0.14)
\end{equation}

These results highlight the strong interplay between X-ray output and the black hole mass, reinforcing the connection between AGN central activity and its observable properties across optical and X-ray wavelengths.

\section{Discussion}
\label{sec:discussion}

\subsection{eROSITA and XMM-Newton samples}
\label{sect-D-sample}

A major difference in the approach of defining galaxy samples in this work and in \citep{Pulatova2024} lies in the differences between the XMM-Newton and eROSITA missions. The eROSITA survey covers the entire sky, while XMM-Newton's pointed observations only cover $\approx$3\%\ with a large range in sensitivity. However, the positional accuracy of XMM-Newton source detections is much better than in eROSITA. On the other side, eROSITA observations make it possible to study the broader and more distant samples of the galaxy populations (Fig.~\ref{Fig-z}).
Our sample benefits from the SDSS-V/SPIDERS target list, which employs probabilistic associations.  The adopted selection method—based on NWAY and machine-learning classification—optimally balances positional accuracy with multiwavelength source properties, achieving high completeness and reliability.

For faint sources, eROSITA X-ray fluxes may appear systematically higher than those reported by deeper surveys such as XMM-Newton, particularly in the soft energy band. This is possibly related to ongoing calibration uncertainties in eRASS1, especially below 1 keV. A detailed comparison and discussion of flux calibration between eROSITA, 4XMM, and Chandra is provided in \cite{Merloni2024A&A}, where these systematic effects are analyzed and quantified.

In our current work, we used only the narrow component of H$\alpha$ emission line, while \citet{Pulatova2024} used total fluxes without decomposition for the BPT. Decomposition systematically results in lower narrow-line fluxes than total fluxes, as it isolates the AGN contribution by removing contamination from the broad-line component and the host galaxy light. This reduction in the H$\alpha$ flux further elevates the X-ray/H$\alpha$ flux ratio in the eROSITA sample.

These differences highlight the importance of consistent methodologies when comparing AGN diagnostics across different datasets.

\subsection{Classification of galaxies with AGN}
\label{sect-D-BPT}

AGN detection can be achieved by several complementary approaches, including diagnostics such as the optical BPT \citep{Baldwinet1981} or WHAN \citep{Cid_Fernandes2011} diagrams, broad-band optical or infrared variability  \citep{Wang2024}, X-ray detection with the luminosity exceeding that from the stellar population \citep{Merloni2014, Fitriana2022}, gamma-ray and radio emission \citep{Dermer2016}. In general, the X-ray selection of AGN is more effective than optical selection because it is less affected by obscuration from intervening material and is less prone to contamination by the light from the host galaxy, which can dilute AGN signals at optical wavelengths \citep{Hickox2018,2025arXiv250213202G}. The BPT diagram uses emission-line ratios ([O\iii]$/\textrm{H}\beta$ vs. [N\ii]$/\textrm{H}\alpha$) to distinguish AGN from star-forming galaxies by identifying ionization mechanisms consistent with AGN. However, this method can be affected by aperture effects, dust attenuation, and the requirement for the detection of emission lines. Therefore, the most reliable candidates for galaxies with active nuclei are obtained when several of the listed criteria are met simultaneously.

In this paper, we applied a strict SNR~$> 3$ criterion for optical emission lines (H$\alpha$, H$\beta$, [O\iii], [N\ii] --- QUALITY > 1), ensuring more reliable results but significantly reducing the sample size. Lowering the S/N threshold allows for more AGN candidates with weak emission lines, which are missing from the current sample. For comparison, \citet{Pucha2024} used SNR~$> 3$ for [O\iii], H$\alpha$, and [N\ii], but a lower threshold of SNR~$> 1$ for H$\beta$, acknowledging its weaker detectability in many galaxies. Similarly, \citet{Ryzhov2024} applied SNR~$> 2$ and noted that increasing the threshold to SNR~$> 3$ had little impact on their conclusions but reduced the number of classified galaxies, particularly those with weak AGN or star-forming activity. This trade-off underscores the importance of carefully selecting SNR criteria to balance between reliability and sample completeness.

Our current study confirms the dependence of a galaxy position on the BPT diagram from its X-ray/H$\alpha$ ratio from \citet{Pulatova2024}. This result agrees with previous studies of the connection between X-ray and optical emission. The logarithm of X-ray-to-optical(continuum) flux ratio $> -1$ was used to identify AGN in their host galaxies \citep{Brandt2005, Fitriana2022}. A correlation has been established between the soft X-ray luminosities and H$\alpha$ luminosities \citep{Elvis1984, Koratkar1995, Halderson2001}. Compared to its significant influence in the optical spectrum, the host galaxy's relatively minor contribution to X-ray emission enhances the reliability of X-ray observations as a method for identifying AGN. 

\subsection{X-ray and optical luminosities, SMBH mass estimation}
\label{sect-D-X-O}

In this analysis, we consider only sources with QUALITY = 2, corresponding to spectra with reliable optical emission-line measurements. While this ensures consistency in the derived trends with X-ray fluxes, it also introduces a selection bias. A substantial number of X-ray–detected sources with lower-quality spectral classifications—many of which may be type I AGN with weak or undetected optical lines—are excluded. These objects could influence the observed flux–flux correlations.

\citet{Georgantopoulos2010} compare the efficiency of X-ray and [O\iii] optical luminosity functions for identifying AGN and discuss the dependence of X-ray and [O\iii] luminosities. The slopes and intercepts obtained are different for different types of Seyfert galaxies, from $0.59$ to $1.04$ and from $-0.17$ (Seyfert 1) to $16.15$, correspondingly. For the X-ray DR19 sample, we obtained a slope of $0.68$ and an intercept of $15$. These numbers agree with data for Seyfert 2 and luminous Seyfert 1 galaxies \citep{Georgantopoulos2010}. In \citet{Panessa2006}, they find a slope and intercept for their $L_X - L_{[OIII]}$ diagram of $1.22$ and $-7.34$, respectively, which differ from our work. 
We attribute the observed scatter to a combination of factors, including sample selection, methodological differences, and intrinsic physical diversity within the AGN population.

The $L_X vs L_{[OIII]}$ relation was also studied in \citet{Agostino2023}, based on a sample of 500 low-redshift AGNs observed by XMM-Newton and SDSS. They showed that so-called "optically dull AGN" are not a distinct class, but instead represent the [O\iii]-underluminous tail of the unimodal $L_X - L_{[OIII]}$ relation. Their analysis further revealed that the degree of [O\iii] underluminosity correlates with host galaxy properties, particularly the specific star formation rate, suggesting that the observed scatter in the relation is influenced by host-driven effects rather than intrinsic AGN differences.

The $M_{\textrm{BH}} - \sigma_{*}$ scaling relation \citep{2000ApJ...539L...9F,2000ApJ...539L..13G}, which links supermassive black hole mass $M_{\textrm{BH}}$ with the velocity dispersion of stars in galaxy bulges, has been a cornerstone of understanding black hole growth and black hole -- host co-evolution \citep{Kormendy2013}. Figure~\ref{Fig_M_BH_LX} shows a close connection between $\sigma_{*}$, $M_{\textrm{BH}}$ and $L_X$, the linear fit of which yielded slope and intercept coefficients $0.57$ and $38.9$, respectively. These values agree with previous research, where a correlation was found between the BH mass and luminosity \citep{Kollmeier2006}. \citet{Panessa2006}, however, observes no direct correlation between $L_X$ and $M_{\textrm{BH}}$ for the sample of 47 local SMBHs, based on the data from Chandra, XMM-Newton, and ASCA observations.

\section{Conclusions}
\label{sec:conclusions}

In this work, we analyzed the X-ray and optical properties of galaxies selected from the SDSS-V/SPIDERS program. The sample is based on the counterparts to eROSITA X-ray sources identified using the NWAY algorithm with machine-learning techniques, as described in \cite{Salvato2022} and \citet{Salvato2025}.

Our main findings and conclusions are:
\begin{enumerate}

\item We created the new sample of X-ray selected galaxies (see Sec.~\ref{subsec:selection}) and applied the {\sc NBursts} technique to analyze 7546 SDSS-V optical spectra of 3684 sources (see Sec.~\ref{subsec:table} and App.~\ref{app:tab} for table description). To ensure the reliability of our results, we applied specific kinematic and S/N criteria to the dataset and examined the properties of the high-quality subsample (QUALITY = 2, see Sec.~\ref{subsec:table}). 

\item e compared the narrow forbidden emission line fluxes
with the published catalogs: RCSED and SDSS-IV DR17. It showed a good agreement
 with SDSS-IV DR17 data and correlation with RCSED (see Fig.~\ref{Fig-line-decom} Sec.~\ref{subsec:emis}).

\item Using QUALITY = 2 subsample we showed that the narrow + broad decomposition of optical emission lines is crucial for the diagnostic BPT diagrams, which rely on the flux ratios of narrow-line emission (e.g., [O\iii]/H$\beta$ and [N\ii]/H$\alpha$) to classify galaxies into star-forming, Seyfert, LINERs, or composite categories.

It leads to a shift in the position on the BPT diagram. We find that 82\% of galaxies move upward along the $\log($[O\iii]/H$\beta)$ axis, with a median shift of +0.17 dex and a 16–84\% percentile range of [+0.07, +0.38] dex.
This upward shift reflects a more precise representation of the AGN ionization mechanism by isolating the narrow-line components less affected by contamination from the broad component and stellar contributions from the host galaxy.

\item We demonstrated a clear and consistent correlation between the X-ray luminosity and the luminosities of both the broad and narrow components of permitted emission lines (H$\alpha$ and H$\beta$), as well as with the narrow forbidden [O\iii] line (see Sec.\ref{subsec:xray}, Figs.\ref{Fig-Lum-nar-broad_X} and~\ref{Fig-lum-[OIII]-Halpha}). The broad-line luminosities trace the immediate vicinity of the SMBH, where gas dynamics are dominated by the central AGN, while the narrow-line components probe ionized gas on larger galactic scales.
These correlations suggest a common physical driver—namely, the AGN's UV accretion disk emission, which powers both the ionized gas and the high-energy X-rays. Although the UV continuum is difficult to observe directly, the slopes and scatter of the optical-X-ray relations offer valuable insights into how this radiation is reprocessed by various AGN structures, and into the role of geometry, covering factor, and ionization state. The nearly identical correlation coefficients found for the broad H$\alpha$ and narrow [O\iii] lines—both with slopes of 0.68 and similar intercepts—indicate that X-ray luminosity scales uniformly with optical line emission, reinforcing its role as a robust tracer of AGN activity across different spatial regions and emission mechanisms.

\item We confirmed the dependence of the position on the BPT diagram as a function of X-ray/H$\alpha$ ratio as reported in \citet{Pulatova2024}, using decomposed narrow-line fluxes, extending the trend to higher X-ray/H$\alpha$ flux ratios ($F_{\textrm{X-ray}}/F_{\textrm{H}\alpha} > 2$) (see Sec.~\ref{subsec:bpt} and Fig.~\ref{Fig-bpt-sdss-v-X-A}).

\item We demonstrate correlations between the position of a source on the BPT diagram and main galaxy parameters (see Fig.~\ref{Fig-bpt-RCSED}), including black hole mass ($M_{\textrm{BH}}$) and velocity dispersion ($\sigma_{\rm narrow}$). Galaxies with higher $M_{\textrm{BH}}$ and $\sigma_{\rm narrow}$ values are systematically located higher on the $\log($[O\iii]/H$\beta)$ axis, indicating a more substantial AGN ionization contribution. This trend underscores the connection between the central black hole's properties and the ionization state of the narrow-line region.

In future work, we plan to explore several open questions in more detail. For example, we aim to apply completeness corrections based on the full SDSS-V selection function and to model selection biases more quantitatively using mock catalogs. A more detailed investigation of the physical drivers behind the observed emission-line correlations—such as black hole mass, Eddington ratio, and host galaxy properties—also remains to be done. Additionally, we intend to include diagnostic tools such as the WHAN diagram \citep{Cid_Fernandes2011} to better separate AGN from star formation contributions in the narrow emission lines. These directions will help refine the interpretation of emission-line scaling relations and further constrain the co-evolution of AGN and their host galaxies.
\end{enumerate}

\section*{Data availability}

Table~\ref{app:tab},containing the results presented in this paper, is only available in electronic form at the CDS via anonymous ftp to \url{cdsarc.u-strasbg.fr} (130.79.128.5) or via \url{http://cdsweb.u-strasbg.fr/cgi-bin/qcat?J/A+A/}
A detailed description of the table columns is provided in Table~\ref{app:tab}.
The posteriors for LX-LOIII scaling relation are available on \href{https://zenodo.org/records/16520014}{Zenodo}~\citep{zenodo_lxlo}.

\begin{acknowledgements}
Funding for the Sloan Digital Sky Survey V (SDSS-V) has been provided by the Alfred P. Sloan Foundation, the Heising-Simons Foundation, the National Science Foundation, and the participating institutions. SDSS-V is managed by the Astrophysical Research Consortium for the Participating Institutions of the SDSS Collaboration. For a full list of SDSS-V participating institutions, please see \href{https://www.sdss.org/}{https://www.sdss.org/} We are grateful to all contributors who have enabled SDSS-V's scientific capabilities. Part of this work was supported by the German Deut\-sche For\-schungs\-ge\-mein\-schaft, DFG\/, Water Benjamin Stelle, project number PU~848/1--1. Igor Chilingarian's research is supported by the Telescope Data Center at the Smithsonian Astrophysical Observatory. He also acknowledges the support from the NASA ADAP-22-0102 grant (award 80NSSC23K0493). ER and DG's research on the analysis and fitting of spectra was supported by the Russian Science Foundation (RScF) grant No.~23-12-00146. C. Aydar acknowledges the support by the Excellence Cluster ORIGINS, which is funded by the Deutsche Forschungsgemeinschaft (DFG, German Research Foundation) under Germany's Excellence Strategy - EXC-2094 - 390783311. MD expresses gratitude to the International Max Planck Research School for Astronomy and Cosmic Physics at the University of Heidelberg (IMPRS-HD). Nadiia Pulatova is grateful for the scientific discussion to Anatolii Tugay, Lidiia Zadorozhna; and for IT support to Stefan Kallweit; and for inspiration to Emiliia Pulatova. RJA was supported by FONDECYT grant number 1231718 and by the ANID BASAL project FB210003. FEB acknowledges support from ANID-Chile BASAL CATA FB210003, FONDECYT Regular 1241005, and Millennium Science Initiative, AIM23-0001. CAN and HJIM acknowledge the support from projects CONAHCyT CBF2023-2024-1418, PAPIIT IA104325 and IN119123. This research made use of TOPCAT, a software tool for astronomical data analysis \citep{Taylor2005}. We acknowledge Mark Taylor for developing and maintaining this valuable resource. We acknowledge the use of ChatGPT, Grammarly, and DeepL for language refinement and improving the manuscript's clarity. The final content remains the responsibility of the authors. We complement the use of LLMs with critical thinking, as was proposed in \citep{2024arXiv240920252F}.
\end{acknowledgements}

\bibliographystyle{aa} 
\bibliography{aa55117-25} 

\begin{thebibliography}{86}
\expandafter\ifx\csname natexlab\endcsname\relax\def\natexlab#1{#1}\fi

\bibitem[{{Abdurro'uf} {et~al.}(2022){Abdurro'uf}, {Accetta}, {Aerts}, {Silva Aguirre}, {Ahumada}, {Ajgaonkar}, {Filiz Ak}, {Alam}, {Allende Prieto}, {Almeida}, {Anders}, {Anderson}, {Andrews}, {Anguiano}, {Aquino-Ort{\'\i}z}, {Arag{\'o}n-Salamanca}, {Argudo-Fern{\'a}ndez}, {Ata}, {Aubert}, {Avila-Reese}, {Badenes}, {Barb{\'a}}, {Barger}, {Barrera-Ballesteros}, {Beaton}, {Beers}, {Belfiore}, {Bender}, {Bernardi}, {Bershady}, {Beutler}, {Bidin}, {Bird}, {Bizyaev}, {Blanc}, {Blanton}, {Boardman}, {Bolton}, {Boquien}, {Borissova}, {Bovy}, {Brandt}, {Brown}, {Brownstein}, {Brusa}, {Buchner}, {Bundy}, {Burchett}, {Bureau}, {Burgasser}, {Cabang}, {Campbell}, {Cappellari}, {Carlberg}, {Wanderley}, {Carrera}, {Cash}, {Chen}, {Chen}, {Cherinka}, {Chiappini}, {Choi}, {Chojnowski}, {Chung}, {Clerc}, {Cohen}, {Comerford}, {Comparat}, {da Costa}, {Covey}, {Crane}, {Cruz-Gonzalez}, {Culhane}, {Cunha}, {Dai}, {Damke}, {Darling}, {Davidson}, {Davies}, {Dawson}, {De Lee}, {Diamond-Stanic}, {Cano-D{\'\i}az}, {S{\'a}nchez},
  {Donor}, {Duckworth}, {Dwelly}, {Eisenstein}, {Elsworth}, {Emsellem}, {Eracleous}, {Escoffier}, {Fan}, {Farr}, {Feng}, {Fern{\'a}ndez-Trincado}, {Feuillet}, {Filipp}, {Fillingham}, {Frinchaboy}, {Fromenteau}, {Galbany}, {Garc{\'\i}a}, {Garc{\'\i}a-Hern{\'a}ndez}, {Ge}, {Geisler}, {Gelfand}, {G{\'e}ron}, {Gibson}, {Goddy}, {Godoy-Rivera}, {Grabowski}, {Green}, {Greener}, {Grier}, {Griffith}, {Guo}, {Guy}, {Hadjara}, {Harding}, {Hasselquist}, {Hayes}, {Hearty}, {Hern{\'a}ndez}, {Hill}, {Hogg}, {Holtzman}, {Horta}, {Hsieh}, {Hsu}, {Hsu}, {Huber}, {Huertas-Company}, {Hutchinson}, {Hwang}, {Ibarra-Medel}, {Chitham}, {Ilha}, {Imig}, {Jaekle}, {Jayasinghe}, {Ji}, {Johnson}, {Jones}, {J{\"o}nsson}, {Katkov}, {Khalatyan}, {Kinemuchi}, {Kisku}, {Knapen}, {Kneib}, {Kollmeier}, {Kong}, {Kounkel}, {Kreckel}, {Krishnarao}, {Lacerna}, {Lane}, {Langgin}, {Lavender}, {Law}, {Lazarz}, {Leung}, {Leung}, {Lewis}, {Li}, {Li}, {Lian}, {Liang}, {Lin}, {Lin}, {Lin}, {Lintott}, {Long}, {Longa-Pe{\~n}a}, {L{\'o}pez-Cob{\'a}}, {Lu},
  {Lundgren}, {Luo}, {Mackereth}, {de la Macorra}, {Mahadevan}, {Majewski}, {Manchado}, {Mandeville}, {Maraston}, {Margalef-Bentabol}, {Masseron}, {Masters}, {Mathur}, {McDermid}, {Mckay}, {Merloni}, {Merrifield}, {Meszaros}, {Miglio}, {Di Mille}, {Minniti}, {Minsley}, {Monachesi}, {Moon}, {Mosser}, {Mulchaey}, {Muna}, {Mu{\~n}oz}, {Myers}, {Myers}, {Nadathur}, {Nair}, {Nandra}, {Neumann}, {Newman}, {Nidever}, {Nikakhtar}, {Nitschelm}, {O'Connell}, {Garma-Oehmichen}, {Luan Souza de Oliveira}, {Olney}, {Oravetz}, {Ortigoza-Urdaneta}, {Osorio}, {Otter}, {Pace}, {Padilla}, {Pan}, {Pan}, {Parikh}, {Parker}, {Peirani}, {Pe{\~n}a Ram{\'\i}rez}, {Penny}, {Percival}, {Perez-Fournon}, {Pinsonneault}, {Poidevin}, {Poovelil}, {Price-Whelan}, {B{\'a}rbara de Andrade Queiroz}, {Raddick}, {Ray}, {Rembold}, {Riddle}, {Riffel}, {Riffel}, {Rix}, {Robin}, {Rodr{\'\i}guez-Puebla}, {Roman-Lopes}, {Rom{\'a}n-Z{\'u}{\~n}iga}, {Rose}, {Ross}, {Rossi}, {Rubin}, {Salvato}, {S{\'a}nchez}, {S{\'a}nchez-Gallego}, {Sanderson}, {Santana
  Rojas}, {Sarceno}, {Sarmiento}, {Sayres}, {Sazonova}, {Schaefer}, {Schiavon}, {Schlegel}, {Schneider}, {Schultheis}, {Schwope}, {Serenelli}, {Serna}, {Shao}, {Shapiro}, {Sharma}, {Shen}, {Shetrone}, {Shu}, {Simon}, {Skrutskie}, {Smethurst}, {Smith}, {Sobeck}, {Spoo}, {Sprague}, {Stark}, {Stassun}, {Steinmetz}, {Stello}, {Stone-Martinez}, {Storchi-Bergmann}, {Stringfellow}, {Stutz}, {Su}, {Taghizadeh-Popp}, {Talbot}, {Tayar}, {Telles}, {Teske}, {Thakar}, {Theissen}, {Tkachenko}, {Thomas}, {Tojeiro}, {Hernandez Toledo}, {Troup}, {Trump}, {Trussler}, {Turner}, {Tuttle}, {Unda-Sanzana}, {V{\'a}zquez-Mata}, {Valentini}, {Valenzuela}, {Vargas-Gonz{\'a}lez}, {Vargas-Maga{\~n}a}, {Alfaro}, {Villanova}, {Vincenzo}, {Wake}, {Warfield}, {Washington}, {Weaver}, {Weijmans}, {Weinberg}, {Weiss}, {Westfall}, {Wild}, {Wilde}, {Wilson}, {Wilson}, {Wilson}, {Wolf}, {Wood-Vasey}, {Yan}, {Zamora}, {Zasowski}, {Zhang}, {Zhao}, {Zheng}, {Zheng}, \& {Zhu}}]{Abdurro'uf2022}
{Abdurro'uf}, {Accetta}, K., {Aerts}, C., {et~al.} 2022, \apjs, 259, 35

\bibitem[{{Agostino} {et~al.}(2023){Agostino}, {Salim}, {Ellison}, {Bickley}, \& {Faber}}]{Agostino2023}
{Agostino}, C.~J., {Salim}, S., {Ellison}, S.~L., {Bickley}, R.~W., \& {Faber}, S.~M. 2023, \apj, 943, 174

\bibitem[{{Allen} {et~al.}(2008){Allen}, {Groves}, {Dopita}, {Sutherland}, \& {Kewley}}]{Allen2008}
{Allen}, M.~G., {Groves}, B.~A., {Dopita}, M.~A., {Sutherland}, R.~S., \& {Kewley}, L.~J. 2008, \apjs, 178, 20

\bibitem[{{Antonucci}(1993)}]{Antonucci1993}
{Antonucci}, R. 1993, \araa, 31, 473

\bibitem[{{Aydar} {et~al.}(2025){Aydar}, {Merloni}, {Dwelly}, {Comparat}, {Salvato}, {Buchner}, {Brusa}, {Liu}, {Wolf}, {Anderson}, {Andonie}, {Bauer}, {Blanton}, {Brandt}, {D{\'\i}az}, {Hern{\'a}ndez-Garc{\'\i}a}, {Kim}, {Miyaji}, {Morrison}, {Musiimenta}, {Negrete}, {Ni}, {Ricci}, {Schneider}, {Schwope}, {Shen}, {Waddell}, {Arcodia}, {Bizyaev}, {Burchett}, {Chakraborty}, {Covey}, {G{\"a}nsicke}, {Georgakakis}, {Green}, {Ibarra}, {Ider-Chitham}, {Koekemoer}, {Kollmeier}, {Krumpe}, {Lamer}, {Malyali}, {Nandra}, {Pan}, {Pizarro}, {S{\'a}nchez-Gallego}, {Trump}, \& {Urrutia}}]{2025A&A...698A.132A}
{Aydar}, C., {Merloni}, A., {Dwelly}, T., {et~al.} 2025, \aap, 698, A132

\bibitem[{{Baldwin} {et~al.}(1981){Baldwin}, {Phillips}, \& {Terlevich}}]{Baldwinet1981}
{Baldwin}, J.~A., {Phillips}, M.~M., \& {Terlevich}, R. 1981, \pasp, 93, 5

\bibitem[{{Bentz} {et~al.}(2013){Bentz}, {Denney}, {Grier}, {Barth}, {Peterson}, {Vestergaard}, {Bennert}, {Canalizo}, {De Rosa}, {Filippenko}, {Gates}, {Greene}, {Li}, {Malkan}, {Pogge}, {Stern}, {Treu}, \& {Woo}}]{2013ApJ...767..149B}
{Bentz}, M.~C., {Denney}, K.~D., {Grier}, C.~J., {et~al.} 2013, \apj, 767, 149

\bibitem[{{Blanton} {et~al.}(2017){Blanton}, {Bershady}, {Abolfathi}, {Albareti}, {Allende Prieto}, {Almeida}, {Alonso-Garc{\'\i}a}, {Anders}, {Anderson}, {Andrews}, {Aquino-Ort{\'\i}z}, {Arag{\'o}n-Salamanca}, {Argudo-Fern{\'a}ndez}, {Armengaud}, {Aubourg}, {Avila-Reese}, {Badenes}, {Bailey}, {Barger}, {Barrera-Ballesteros}, {Bartosz}, {Bates}, {Baumgarten}, {Bautista}, {Beaton}, {Beers}, {Belfiore}, {Bender}, {Berlind}, {Bernardi}, {Beutler}, {Bird}, {Bizyaev}, {Blanc}, {Blomqvist}, {Bolton}, {Boquien}, {Borissova}, {van den Bosch}, {Bovy}, {Brandt}, {Brinkmann}, {Brownstein}, {Bundy}, {Burgasser}, {Burtin}, {Busca}, {Cappellari}, {Delgado Carigi}, {Carlberg}, {Carnero Rosell}, {Carrera}, {Chanover}, {Cherinka}, {Cheung}, {G{\'o}mez Maqueo Chew}, {Chiappini}, {Choi}, {Chojnowski}, {Chuang}, {Chung}, {Cirolini}, {Clerc}, {Cohen}, {Comparat}, {da Costa}, {Cousinou}, {Covey}, {Crane}, {Croft}, {Cruz-Gonzalez}, {Garrido Cuadra}, {Cunha}, {Damke}, {Darling}, {Davies}, {Dawson}, {de la Macorra}, {Dell'Agli}, {De
  Lee}, {Delubac}, {Di Mille}, {Diamond-Stanic}, {Cano-D{\'\i}az}, {Donor}, {Downes}, {Drory}, {du Mas des Bourboux}, {Duckworth}, {Dwelly}, {Dyer}, {Ebelke}, {Eigenbrot}, {Eisenstein}, {Emsellem}, {Eracleous}, {Escoffier}, {Evans}, {Fan}, {Fern{\'a}ndez-Alvar}, {Fernandez-Trincado}, {Feuillet}, {Finoguenov}, {Fleming}, {Font-Ribera}, {Fredrickson}, {Freischlad}, {Frinchaboy}, {Fuentes}, {Galbany}, {Garcia-Dias}, {Garc{\'\i}a-Hern{\'a}ndez}, {Gaulme}, {Geisler}, {Gelfand}, {Gil-Mar{\'\i}n}, {Gillespie}, {Goddard}, {Gonzalez-Perez}, {Grabowski}, {Green}, {Grier}, {Gunn}, {Guo}, {Guy}, {Hagen}, {Hahn}, {Hall}, {Harding}, {Hasselquist}, {Hawley}, {Hearty}, {Gonzalez Hern{\'a}ndez}, {Ho}, {Hogg}, {Holley-Bockelmann}, {Holtzman}, {Holzer}, {Huehnerhoff}, {Hutchinson}, {Hwang}, {Ibarra-Medel}, {da Silva Ilha}, {Ivans}, {Ivory}, {Jackson}, {Jensen}, {Johnson}, {Jones}, {J{\"o}nsson}, {Jullo}, {Kamble}, {Kinemuchi}, {Kirkby}, {Kitaura}, {Klaene}, {Knapp}, {Kneib}, {Kollmeier}, {Lacerna}, {Lane}, {Lang}, {Law},
  {Lazarz}, {Lee}, {Le Goff}, {Liang}, {Li}, {Li}, {Lian}, {Lima}, {Lin}, {Lin}, {Bertran de Lis}, {Liu}, {de Icaza Lizaola}, {Long}, {Lucatello}, {Lundgren}, {MacDonald}, {Deconto Machado}, {MacLeod}, {Mahadevan}, {Geimba Maia}, {Maiolino}, {Majewski}, {Malanushenko}, {Malanushenko}, {Manchado}, {Mao}, {Maraston}, {Marques-Chaves}, {Masseron}, {Masters}, {McBride}, {McDermid}, {McGrath}, {McGreer}, {Medina Pe{\~n}a}, {Melendez}, {Merloni}, {Merrifield}, {Meszaros}, {Meza}, {Minchev}, {Minniti}, {Miyaji}, {More}, {Mulchaey}, {M{\"u}ller-S{\'a}nchez}, {Muna}, {Munoz}, {Myers}, {Nair}, {Nandra}, {Correa do Nascimento}, {Negrete}, {Ness}, {Newman}, {Nichol}, {Nidever}, {Nitschelm}, {Ntelis}, {O'Connell}, {Oelkers}, {Oravetz}, {Oravetz}, {Pace}, {Padilla}, {Palanque-Delabrouille}, {Alonso Palicio}, {Pan}, {Parejko}, {Parikh}, {P{\^a}ris}, {Park}, {Patten}, {Peirani}, {Pellejero-Ibanez}, {Penny}, {Percival}, {Perez-Fournon}, {Petitjean}, {Pieri}, {Pinsonneault}, {Pisani}, {Poleski}, {Prada}, {Prakash}, {Queiroz},
  {Raddick}, {Raichoor}, {Barboza Rembold}, {Richstein}, {Riffel}, {Riffel}, {Rix}, {Robin}, {Rockosi}, {Rodr{\'\i}guez-Torres}, {Roman-Lopes}, {Rom{\'a}n-Z{\'u}{\~n}iga}, {Rosado}, {Ross}, {Rossi}, {Ruan}, {Ruggeri}, {Rykoff}, {Salazar-Albornoz}, {Salvato}, {S{\'a}nchez}, {Aguado}, {S{\'a}nchez-Gallego}, {Santana}, {Santiago}, {Sayres}, {Schiavon}, {da Silva Schimoia}, {Schlafly}, {Schlegel}, {Schneider}, {Schultheis}, {Schuster}, {Schwope}, {Seo}, {Shao}, {Shen}, {Shetrone}, {Shull}, {Simon}, {Skinner}, {Skrutskie}, {Slosar}, {Smith}, {Sobeck}, {Sobreira}, {Somers}, {Souto}, {Stark}, {Stassun}, {Stauffer}, {Steinmetz}, {Storchi-Bergmann}, {Streblyanska}, {Stringfellow}, {Su{\'a}rez}, {Sun}, {Suzuki}, {Szigeti}, {Taghizadeh-Popp}, {Tang}, {Tao}, {Tayar}, {Tembe}, {Teske}, {Thakar}, {Thomas}, {Thompson}, {Tinker}, {Tissera}, {Tojeiro}, {Hernandez Toledo}, {de la Torre}, {Tremonti}, {Troup}, {Valenzuela}, {Martinez Valpuesta}, {Vargas-Gonz{\'a}lez}, {Vargas-Maga{\~n}a}, {Vazquez}, {Villanova}, {Vivek}, {Vogt},
  {Wake}, {Walterbos}, {Wang}, {Weaver}, {Weijmans}, {Weinberg}, {Westfall}, {Whelan}, {Wild}, {Wilson}, {Wood-Vasey}, {Wylezalek}, {Xiao}, {Yan}, {Yang}, {Ybarra}, {Y{\`e}che}, {Zakamska}, {Zamora}, {Zarrouk}, {Zasowski}, {Zhang}, {Zhao}, {Zheng}, {Zheng}, {Zhou}, {Zhou}, {Zhu}, {Zoccali}, \& {Zou}}]{Blanton2017}
{Blanton}, M.~R., {Bershady}, M.~A., {Abolfathi}, B., {et~al.} 2017, \aj, 154, 28

\bibitem[{{Brandt} \& {Hasinger}(2005)}]{Brandt2005}
{Brandt}, W.~N. \& {Hasinger}, G. 2005, \araa, 43, 827

\bibitem[{{Buchner}(2021)}]{2021JOSS....6.3001B}
{Buchner}, J. 2021, The Journal of Open Source Software, 6, 3001

\bibitem[{{Buchner}(2023)}]{2023StSur..17..169B}
{Buchner}, J. 2023, Statistics Surveys, 17, 169

\bibitem[{{Cappellari}(2017)}]{2017MNRAS.466..798C}
{Cappellari}, M. 2017, \mnras, 466, 798

\bibitem[{{Chilingarian} {et~al.}(2024){Chilingarian}, {Borisov}, {Goradzhanov}, {Grishin}, {Kasparova}, {Katkov}, {Klochkov}, {Rubtsov}, \& {Toptun}}]{2024ASPC..535..179C}
{Chilingarian}, I., {Borisov}, S., {Goradzhanov}, V., {et~al.} 2024, in Astronomical Society of the Pacific Conference Series, Vol. 535, Astromical Data Analysis Software and Systems XXXI, ed. B.~V. {Hugo}, R.~{Van Rooyen}, \& O.~M. {Smirnov}, 179

\bibitem[{{Chilingarian} {et~al.}(2007{\natexlab{a}}){Chilingarian}, {Prugniel}, {Sil'chenko}, \& {Koleva}}]{2007IAUS..241..175C}
{Chilingarian}, I., {Prugniel}, P., {Sil'chenko}, O., \& {Koleva}, M. 2007{\natexlab{a}}, in IAU Symposium, Vol. 241, Stellar Populations as Building Blocks of Galaxies, ed. A.~r. {Vazdekis} \& R.~{Peletier}, 175--176

\bibitem[{{Chilingarian} \& {Asa'd}(2018)}]{2018ApJ...858...63C}
{Chilingarian}, I.~V. \& {Asa'd}, R.~a. 2018, \apj, 858, 63

\bibitem[{{Chilingarian} {et~al.}(2018){Chilingarian}, {Katkov}, {Zolotukhin}, {Grishin}, {Beletsky}, {Boutsia}, \& {Osip}}]{2018ApJ...863....1C}
{Chilingarian}, I.~V., {Katkov}, I.~Y., {Zolotukhin}, I.~Y., {et~al.} 2018, \apj, 863, 1

\bibitem[{{Chilingarian} {et~al.}(2007{\natexlab{b}}){Chilingarian}, {Prugniel}, {Sil'Chenko}, \& {Afanasiev}}]{2007MNRAS.376.1033C}
{Chilingarian}, I.~V., {Prugniel}, P., {Sil'Chenko}, O.~K., \& {Afanasiev}, V.~L. 2007{\natexlab{b}}, \mnras, 376, 1033

\bibitem[{{Chilingarian} {et~al.}(2017){Chilingarian}, {Zolotukhin}, {Katkov}, {Melchior}, {Rubtsov}, \& {Grishin}}]{Chilingarian2017}
{Chilingarian}, I.~V., {Zolotukhin}, I.~Y., {Katkov}, I.~Y., {et~al.} 2017, \apjs, 228, 14

\bibitem[{{Cid Fernandes} {et~al.}(2011){Cid Fernandes}, {Stasi{\'n}ska}, {Mateus}, \& {Vale Asari}}]{Cid_Fernandes2011}
{Cid Fernandes}, R., {Stasi{\'n}ska}, G., {Mateus}, A., \& {Vale Asari}, N. 2011, \mnras, 413, 1687

\bibitem[{Collaboration {et~al.}(2025)Collaboration, Pallathadka, Aghakhanloo, Aird, Almeida, Amrita, Anders, Anderson, Arseneau, Avila, Aviram, Aydar, Badenes, Barrera-Ballesteros, Bauer, Behmard, Berg, Besser, Bidin, Bizyaev, Blanc, Blanton, Bovy, Brandt, Brownstein, Buchner, Bulbul, Burchett, Carigi, Carlberg, Casey, Chakraborty, Chanamé, Chandra, Chiappini, Chilingarian, Comparat, Covey, Crumpler, Cunha, D'Onghia, Dai, Darling, Davis, Lee, Deacon, Delgado, Demasi, Demianenko, Demke, Donor, Drory, Durango, Dwelly, Egorov, Egorova, El-Badry, Eracleous, Fan, Farr, Finkbeiner, Fries, Frinchaboy, Fusillo, Félix, Gaensicke, Galligan, García, Gelfand, Grabowski, Grebel, Green, Greve, Grier, Griffith, Guetzoyan, Gupta, Hackshaw, Hall, Hawkins, Hegedűs, Hekker, Herbst, Hermes, Hernández-García, Hiremath, Hogg, Holtzman, Horne, Horta, Huang, Hutchinson, Häberle, Ibarra-Medel, Ji, Jofre, Johnson, Johnson, Johnston, Kaldor, Katkov, Khalatyan, Khoperskov, Klessen, Kluge, Koekemoer, Kollmeier, Kounkel, Kreckel,
  Krishnarao, Krumpe, Lacerna, Laporte, Lepine, Li, Liang, Limberg, Liu, Loebman, Long, Lu, Lucey, Lugo-Aranda, Martinez-Aldama, McKinnon, Medan, Merloni, Morrison, Myers, Mészáros, Müller-Horn, Nepal, Ness, Nidever, Nitschelm, Oravetz, Otto, Pan, Paolino, Peñaloza, Pinsonneault, Popp, Price-Whelan, Pulatova, Queiroz, Raddick, Rankine, Rix, Román-Zúñiga, Rosso, Runnoe, Saad, Salvato, Sanchez, Sattler, Saydjari, Sayres, Schlaufman, Schneider, Schwope, Seaton, Seeburger, Serna, Sharma, Shen, Sinha, Sizemore, Sniegowska, Song, Souto, Stassun, Steinmetz, Stone, Stone-Martinez, Stringfellow, Sánchez, Sánchez-Gallego, Tan, Tayar, Thai, Thakar, Thibodeaux, Ting, Tkachenko, Trakhtenbrot, Trincado, Troup, Trump, Ulloa, der Marel, Vera, Villanova, Villaseñor, Wang, Way, Weijmans, Wheeler, Wilson, Wofford, Wong, Wu, Wylezalek, Xue, Yan, Yang, Zakamska, Zari, Zasowski, Zeltyn, Zheng, Zucker, \& de~J.~Zermeño}]{SDSS_Collaboration2025}
Collaboration, S., Pallathadka, G.~A., Aghakhanloo, M., {et~al.} 2025, The Nineteenth Data Release of the Sloan Digital Sky Survey

\bibitem[{{Comparat} {et~al.}(2019){Comparat}, {Merloni}, {Salvato}, {Nandra}, {Boller}, {Georgakakis}, {Finoguenov}, {Dwelly}, {Buchner}, {Del Moro}, {Clerc}, {Wang}, {Zhao}, {Prada}, {Yepes}, {Brusa}, {Krumpe}, \& {Liu}}]{Comparat2019}
{Comparat}, J., {Merloni}, A., {Salvato}, M., {et~al.} 2019, \mnras, 487, 2005

\bibitem[{{Cortes-Su{\'a}rez} {et~al.}(2022){Cortes-Su{\'a}rez}, {Negrete}, {Hern{\'a}ndez-Toledo}, {Ibarra-Medel}, \& {Lacerna}}]{Cortes-Suarez2022}
{Cortes-Su{\'a}rez}, E., {Negrete}, C.~A., {Hern{\'a}ndez-Toledo}, H.~M., {Ibarra-Medel}, H., \& {Lacerna}, I. 2022, \mnras, 514, 3626

\bibitem[{Demianenko \& Pulatova(2025)}]{zenodo_lxlo}
Demianenko, M. \& Pulatova, N. 2025, Posteriors for LX-LOIII

\bibitem[{{Dermer} \& {Giebels}(2016)}]{Dermer2016}
{Dermer}, C.~D. \& {Giebels}, B. 2016, Comptes Rendus Physique, 17, 594

\bibitem[{{Dey} {et~al.}(2019){Dey}, {Schlegel}, {Lang}, {Blum}, {Burleigh}, {Fan}, {Findlay}, {Finkbeiner}, {Herrera}, {Juneau}, {Landriau}, {Levi}, {McGreer}, {Meisner}, {Myers}, {Moustakas}, {Nugent}, {Patej}, {Schlafly}, {Walker}, {Valdes}, {Weaver}, {Y{\`e}che}, {Zou}, {Zhou}, {Abareshi}, {Abbott}, {Abolfathi}, {Aguilera}, {Alam}, {Allen}, {Alvarez}, {Annis}, {Ansarinejad}, {Aubert}, {Beechert}, {Bell}, {BenZvi}, {Beutler}, {Bielby}, {Bolton}, {Brice{\~n}o}, {Buckley-Geer}, {Butler}, {Calamida}, {Carlberg}, {Carter}, {Casas}, {Castander}, {Choi}, {Comparat}, {Cukanovaite}, {Delubac}, {DeVries}, {Dey}, {Dhungana}, {Dickinson}, {Ding}, {Donaldson}, {Duan}, {Duckworth}, {Eftekharzadeh}, {Eisenstein}, {Etourneau}, {Fagrelius}, {Farihi}, {Fitzpatrick}, {Font-Ribera}, {Fulmer}, {G{\"a}nsicke}, {Gaztanaga}, {George}, {Gerdes}, {Gontcho}, {Gorgoni}, {Green}, {Guy}, {Harmer}, {Hernandez}, {Honscheid}, {Huang}, {James}, {Jannuzi}, {Jiang}, {Joyce}, {Karcher}, {Karkar}, {Kehoe}, {Kneib}, {Kueter-Young}, {Lan},
  {Lauer}, {Le Guillou}, {Le Van Suu}, {Lee}, {Lesser}, {Perreault Levasseur}, {Li}, {Mann}, {Marshall}, {Mart{\'\i}nez-V{\'a}zquez}, {Martini}, {du Mas des Bourboux}, {McManus}, {Meier}, {M{\'e}nard}, {Metcalfe}, {Mu{\~n}oz-Guti{\'e}rrez}, {Najita}, {Napier}, {Narayan}, {Newman}, {Nie}, {Nord}, {Norman}, {Olsen}, {Paat}, {Palanque-Delabrouille}, {Peng}, {Poppett}, {Poremba}, {Prakash}, {Rabinowitz}, {Raichoor}, {Rezaie}, {Robertson}, {Roe}, {Ross}, {Ross}, {Rudnick}, {Safonova}, {Saha}, {S{\'a}nchez}, {Savary}, {Schweiker}, {Scott}, {Seo}, {Shan}, {Silva}, {Slepian}, {Soto}, {Sprayberry}, {Staten}, {Stillman}, {Stupak}, {Summers}, {Sien Tie}, {Tirado}, {Vargas-Maga{\~n}a}, {Vivas}, {Wechsler}, {Williams}, {Yang}, {Yang}, {Yapici}, {Zaritsky}, {Zenteno}, {Zhang}, {Zhang}, {Zhou}, \& {Zhou}}]{Dey2019}
{Dey}, A., {Schlegel}, D.~J., {Lang}, D., {et~al.} 2019, \aj, 157, 168

\bibitem[{{Dwelly} {et~al.}(2017){Dwelly}, {Salvato}, {Merloni}, {Brusa}, {Buchner}, {Anderson}, {Boller}, {Brandt}, {Budav{\'a}ri}, {Clerc}, {Coffey}, {Del Moro}, {Georgakakis}, {Green}, {Jin}, {Menzel}, {Myers}, {Nandra}, {Nichol}, {Ridl}, {Schwope}, \& {Simm}}]{Dwelly2017}
{Dwelly}, T., {Salvato}, M., {Merloni}, A., {et~al.} 2017, \mnras, 469, 1065

\bibitem[{{Elvis} {et~al.}(1984){Elvis}, {Soltan}, \& {Keel}}]{Elvis1984}
{Elvis}, M., {Soltan}, A., \& {Keel}, W.~C. 1984, \apj, 283, 479

\bibitem[{{Fabian}(2012)}]{Fabian2012}
{Fabian}, A.~C. 2012, \araa, 50, 455

\bibitem[{{Ferland} \& {Netzer}(1983)}]{Ferland1983}
{Ferland}, G.~J. \& {Netzer}, H. 1983, \apj, 264, 105

\bibitem[{{Ferrarese} \& {Merritt}(2000)}]{2000ApJ...539L...9F}
{Ferrarese}, L. \& {Merritt}, D. 2000, \apjl, 539, L9

\bibitem[{{Fitriana} \& {Murayama}(2022)}]{Fitriana2022}
{Fitriana}, I.~K. \& {Murayama}, T. 2022, \pasj, 74, 689

\bibitem[{{Fouesneau} {et~al.}(2024){Fouesneau}, {Momcheva}, {Chadayammuri}, {Demianenko}, {Dumont}, {Hviding}, {Kahle}, {Pulatova}, {Rajpoot}, {Scheuck}, {Seeburger}, {Semenov}, \& {Villase{\~n}or}}]{2024arXiv240920252F}
{Fouesneau}, M., {Momcheva}, I.~G., {Chadayammuri}, U., {et~al.} 2024, arXiv e-prints, arXiv:2409.20252

\bibitem[{{Gebhardt} {et~al.}(2000){Gebhardt}, {Bender}, {Bower}, {Dressler}, {Faber}, {Filippenko}, {Green}, {Grillmair}, {Ho}, {Kormendy}, {Lauer}, {Magorrian}, {Pinkney}, {Richstone}, \& {Tremaine}}]{2000ApJ...539L..13G}
{Gebhardt}, K., {Bender}, R., {Bower}, G., {et~al.} 2000, \apjl, 539, L13

\bibitem[{{Georgantopoulos} \& {Akylas}(2010)}]{Georgantopoulos2010}
{Georgantopoulos}, I. \& {Akylas}, A. 2010, \aap, 509, A38

\bibitem[{{Giacconi} {et~al.}(1962){Giacconi}, {Gursky}, {Paolini}, \& {Rossi}}]{Giacconi1962}
{Giacconi}, R., {Gursky}, H., {Paolini}, F.~R., \& {Rossi}, B.~B. 1962, \prl, 9, 439

\bibitem[{{Goradzhanov} {et~al.}(2024){Goradzhanov}, {Chilingarian}, {Rubtsov}, {Katkov}, {Grishin}, {Toptun}, {Kasparova}, {Klochkov}, \& {Borisov}}]{2024ASPC..535..175G}
{Goradzhanov}, V., {Chilingarian}, I., {Rubtsov}, E., {et~al.} 2024, in Astronomical Society of the Pacific Conference Series, Vol. 535, Astromical Data Analysis Software and Systems XXXI, ed. B.~V. {Hugo}, R.~{Van Rooyen}, \& O.~M. {Smirnov}, 175

\bibitem[{{Greene} \& {Ho}(2005)}]{2005ApJ...630..122G}
{Greene}, J.~E. \& {Ho}, L.~C. 2005, \apj, 630, 122

\bibitem[{{Grishin} {et~al.}(2025){Grishin}, {Chilingarian}, {Combes}, {Bauer}, {Toptun}, {Katkov}, \& {Fabricant}}]{2025arXiv250213202G}
{Grishin}, K.~A., {Chilingarian}, I.~V., {Combes}, F., {et~al.} 2025, arXiv e-prints, arXiv:2502.13202

\bibitem[{{Halderson} {et~al.}(2001){Halderson}, {Moran}, {Filippenko}, \& {Ho}}]{Halderson2001}
{Halderson}, E.~L., {Moran}, E.~C., {Filippenko}, A.~V., \& {Ho}, L.~C. 2001, \aj, 122, 637

\bibitem[{{Harrison} {et~al.}(2012){Harrison}, {Alexander}, {Swinbank}, {Smail}, {Alaghband-Zadeh}, {Bauer}, {Chapman}, {Del Moro}, {Hickox}, {Ivison}, {Men{\'e}ndez-Delmestre}, {Mullaney}, \& {Nesvadba}}]{Harrison2012}
{Harrison}, C.~M., {Alexander}, D.~M., {Swinbank}, A.~M., {et~al.} 2012, \mnras, 426, 1073

\bibitem[{{Heckman} \& {Best}(2014)}]{Heckman2014}
{Heckman}, T.~M. \& {Best}, P.~N. 2014, \araa, 52, 589

\bibitem[{{Hickox} \& {Alexander}(2018)}]{Hickox2018}
{Hickox}, R.~C. \& {Alexander}, D.~M. 2018, \araa, 56, 625

\bibitem[{{Ho}(2008)}]{Ho2008}
{Ho}, L.~C. 2008, \araa, 46, 475

\bibitem[{{Imanishi} \& {Ueno}(1999)}]{Imanishi1999}
{Imanishi}, M. \& {Ueno}, S. 1999, \mnras, 305, 829

\bibitem[{{Kauffmann} {et~al.}(2003){Kauffmann}, {Heckman}, {Tremonti}, {Brinchmann}, {Charlot}, {White}, {Ridgway}, {Brinkmann}, {Fukugita}, {Hall}, {Ivezi{\'c}}, {Richards}, \& {Schneider}}]{Kauffmann2003}
{Kauffmann}, G., {Heckman}, T.~M., {Tremonti}, C., {et~al.} 2003, \mnras, 346, 1055

\bibitem[{{Kewley} \& {Dopita}(2002)}]{Kewley2002}
{Kewley}, L.~J. \& {Dopita}, M.~A. 2002, \apjs, 142, 35

\bibitem[{{Kewley} {et~al.}(2006){Kewley}, {Groves}, {Kauffmann}, \& {Heckman}}]{Kewleyet2006}
{Kewley}, L.~J., {Groves}, B., {Kauffmann}, G., \& {Heckman}, T. 2006, \mnras, 372, 961

\bibitem[{{Klochkov} {et~al.}(2024){Klochkov}, {Katkov}, {Chilingarian}, {Grishin}, {Kasparova}, {Goradzhanov}, {Toptun}, {Rubtsov}, \& {Borisov}}]{2024ASPC..535..243K}
{Klochkov}, V., {Katkov}, I., {Chilingarian}, I., {et~al.} 2024, in Astronomical Society of the Pacific Conference Series, Vol. 535, Astromical Data Analysis Software and Systems XXXI, ed. B.~V. {Hugo}, R.~{Van Rooyen}, \& O.~M. {Smirnov}, 243

\bibitem[{{Koleva} {et~al.}(2008){Koleva}, {Prugniel}, {Ocvirk}, {Le Borgne}, \& {Soubiran}}]{Koleva2008}
{Koleva}, M., {Prugniel}, P., {Ocvirk}, P., {Le Borgne}, D., \& {Soubiran}, C. 2008, \mnras, 385, 1998

\bibitem[{{Kollmeier} {et~al.}(2019){Kollmeier}, {Anderson}, {Blanc}, {Blanton}, {Covey}, {Crane}, {Drory}, {Frinchaboy}, {Froning}, {Johnson}, {Kneib}, {Kreckel}, {Merloni}, {Pellegrini}, {Pogge}, {Ramirez}, {Rix}, {Sayres}, {S{\'a}nchez-Gallego}, {Shen}, {Tkachenko}, {Trump}, {Tuttle}, {Weijmans}, {Zasowski}, {Barbuy}, {Beaton}, {Bergemann}, {Bochanski}, {Brandt}, {Casey}, {Cherinka}, {Eracleous}, {Fan}, {Garc{\'\i}a}, {Green}, {Hekker}, {Lane}, {Longa-Pe{\~n}a}, {Mathur}, {Meza}, {Minchev}, {Myers}, {Nidever}, {Nitschelm}, {O'Connell}, {Price-Whelan}, {Raddick}, {Rossi}, {Sankrit}, {Simon}, {Stutz}, {Ting}, {Trakhtenbrot}, {Weaver}, {Willmer}, \& {Weinberg}}]{Kollmeier2019}
{Kollmeier}, J., {Anderson}, S.~F., {Blanc}, G.~A., {et~al.} 2019, in Bulletin of the American Astronomical Society, Vol.~51, 274

\bibitem[{{Kollmeier} {et~al.}(2006){Kollmeier}, {Onken}, {Kochanek}, {Gould}, {Weinberg}, {Dietrich}, {Cool}, {Dey}, {Eisenstein}, {Jannuzi}, {Le Floc'h}, \& {Stern}}]{Kollmeier2006}
{Kollmeier}, J.~A., {Onken}, C.~A., {Kochanek}, C.~S., {et~al.} 2006, \apj, 648, 128

\bibitem[{Kollmeier {et~al.}(2025)Kollmeier, Rix, Aerts, Aird, Alfaro, Almeida, Anderson, Óscar Jiménez~Arranz, Arseneau, Assef, Aviram, Aydar, Badenes, Bandyopadhyay, Barger, Barkhouser, Bauer, Bender, Besser, Bhattarai, Bilgi, Bird, Bizyaev, Blanc, Blanton, Bochanski, Bovy, Brandon, Brandt, Brownstein, Buchner, Burchett, Carlberg, Casey, Castaneda-Carlos, Chakraborty, Chanamé, Chandra, Cherinka, Chilingarian, Comparat, Cosens, Covey, Crane, Crumpler, Cunha, Cunningham, Dai, Darling, Jr., Davis, Lee, Deacon, Delgado, Demasi, Demianenko, Derwent, D'Onghia, Mille, Dias, Donor, Drory, Dwelly, Egorov, Egorova, El-Badry, Engelman, Eracleous, Fan, Farr, Fries, Frinchaboy, Froning, Gänsicke, García, Gelfand, Fusillo, Glover, Grabowski, Grebel, Green, Grier, Gupta, Gray, Häberle, Hall, Hammond, Hawkins, Harding, Hegedűs, Herbst, Hermes, Hidalgo, Hilder, Hogg, Holtzman, Horta, Huang, Hwang, Ibarra-Medel, Imig, Inight, Jana, Ji, Jofre, Johns, Johnson, Johnson, Johnston, Jones, Katkov, Koekemoer, Kounkel,
  Kreckel, Krishnarao, Krumpe, Kumari, Kupfer, Lacerna, Laporte, Lepine, Li, Liu, Loebman, Long, Roman-Lopes, Lu, Majewski, Maoz, McKinnon, Medan, Merloni, Minniti, Morrison, Myers, Mészáros, Nandra, Nayak, Ness, Nidever, O'Brien, Oeur, Oravetz, Oravetz, Otto, Pallathadka, Palunas, Pan, Pappalardo, Pandey, Peñaloza, Pinsonneault, Pogge, Popp, Price-Whelan, Pulatova, Qiu, Ramirez, Rankine, Ricci, Runnoe, Sanchez, Salvato, Sattler, Saydjari, Sayres, Schlaufman, Schneider, Schreiber, Schwope, Serna, Shen, Sifón, Singh, Sinha, Smee, Song, Souto, Stassun, Steinmetz, Stone-Martinez, Stringfellow, Stutz, José, Sá, nchez Gallego, Tan, Tayar, Thai, Thakar, Ting, Tkachenko, Tovmasian, Trakhtenbrot, Fernández-Trincado, Troup, Trump, Tuttle, van~der Marel, Villanova, Villaseñor, Wachter, Way, Weijmans, Weinberg, Wheeler, Wilson, Wiggins, Wong, Wu, Wylezalek, Xue, Yang, Zakamska, Zari, Zasowski, Zeltyn, Zucker, Zúñiga, \& de~J.~Zermeño}]{Kollmeier2025a}
Kollmeier, J.~A., Rix, H.-W., Aerts, C., {et~al.} 2025, Sloan Digital Sky Survey-V: Pioneering Panoptic Spectroscopy

\bibitem[{{Kollmeier} {et~al.}(2017){Kollmeier}, {Zasowski}, {Rix}, {Johns}, {Anderson}, {Drory}, {Johnson}, {Pogge}, {Bird}, {Blanc}, {Brownstein}, {Crane}, {De Lee}, {Klaene}, {Kreckel}, {MacDonald}, {Merloni}, {Ness}, {O'Brien}, {Sanchez-Gallego}, {Sayres}, {Shen}, {Thakar}, {Tkachenko}, {Aerts}, {Blanton}, {Eisenstein}, {Holtzman}, {Maoz}, {Nandra}, {Rockosi}, {Weinberg}, {Bovy}, {Casey}, {Chaname}, {Clerc}, {Conroy}, {Eracleous}, {G{\"a}nsicke}, {Hekker}, {Horne}, {Kauffmann}, {McQuinn}, {Pellegrini}, {Schinnerer}, {Schlafly}, {Schwope}, {Seibert}, {Teske}, \& {van Saders}}]{Kollmeier2017}
{Kollmeier}, J.~A., {Zasowski}, G., {Rix}, H.-W., {et~al.} 2017, arXiv e-prints, arXiv:1711.03234

\bibitem[{{Koratkar} {et~al.}(1995){Koratkar}, {Deustua}, {Heckman}, {Filippenko}, {Ho}, \& {Rao}}]{Koratkar1995}
{Koratkar}, A., {Deustua}, S.~E., {Heckman}, T., {et~al.} 1995, \apj, 440, 132

\bibitem[{{Kormendy} \& {Ho}(2013)}]{Kormendy2013}
{Kormendy}, J. \& {Ho}, L.~C. 2013, \araa, 51, 511

\bibitem[{{Markwardt}(2009)}]{MPFIT}
{Markwardt}, C.~B. 2009, in Astronomical Society of the Pacific Conference Series, Vol. 411, Astronomical Data Analysis Software and Systems XVIII, ed. D.~A. {Bohlender}, D.~{Durand}, \& P.~{Dowler}, 251

\bibitem[{{Merloni} {et~al.}(2014){Merloni}, {Bongiorno}, {Brusa}, {Iwasawa}, {Mainieri}, {Magnelli}, {Salvato}, {Berta}, {Cappelluti}, {Comastri}, {Fiore}, {Gilli}, {Koekemoer}, {Le Floc'h}, {Lusso}, {Lutz}, {Miyaji}, {Pozzi}, {Riguccini}, {Rosario}, {Silverman}, {Symeonidis}, {Treister}, {Vignali}, \& {Zamorani}}]{Merloni2014}
{Merloni}, A., {Bongiorno}, A., {Brusa}, M., {et~al.} 2014, \mnras, 437, 3550

\bibitem[{{Merloni} {et~al.}(2024{\natexlab{a}}){Merloni}, {Lamer}, {Liu}, {Ramos-Ceja}, {Brunner}, {Bulbul}, {Dennerl}, {Doroshenko}, {Freyberg}, {Friedrich}, {Gatuzz}, {Georgakakis}, {Haberl}, {Igo}, {Kreykenbohm}, {Liu}, {Maitra}, {Malyali}, {Mayer}, {Nandra}, {Predehl}, {Robrade}, {Salvato}, {Sanders}, {Stewart}, {Tub{\'\i}n-Arenas}, {Weber}, {Wilms}, {Arcodia}, {Artis}, {Aschersleben}, {Avakyan}, {Aydar}, {Bahar}, {Balzer}, {Becker}, {Berger}, {Boller}, {Bornemann}, {Br{\"u}ggen}, {Brusa}, {Buchner}, {Burwitz}, {Camilloni}, {Clerc}, {Comparat}, {Coutinho}, {Czesla}, {Dannhauer}, {Dauner}, {Dauser}, {Dietl}, {Dolag}, {Dwelly}, {Egg}, {Ehl}, {Freund}, {Friedrich}, {Gaida}, {Garrel}, {Ghirardini}, {Gokus}, {Gr{\"u}nwald}, {Grandis}, {Grotova}, {Gruen}, {Gueguen}, {H{\"a}mmerich}, {Hamaus}, {Hasinger}, {Haubner}, {Homan}, {Ider Chitham}, {Joseph}, {Joyce}, {K{\"o}nig}, {Kaltenbrunner}, {Khokhriakova}, {Kink}, {Kirsch}, {Kluge}, {Knies}, {Krippendorf}, {Krumpe}, {Kurpas}, {Li}, {Liu}, {Locatelli}, {Lorenz},
  {M{\"u}ller}, {Magaudda}, {Mannes}, {McCall}, {Meidinger}, {Michailidis}, {Migkas}, {Mu{\~n}oz-Giraldo}, {Musiimenta}, {Nguyen-Dang}, {Ni}, {Olechowska}, {Ota}, {Pacaud}, {Pasini}, {Perinati}, {Pires}, {Pommranz}, {Ponti}, {Poppenhaeger}, {P{\"u}hlhofer}, {Rau}, {Reh}, {Reiprich}, {Roster}, {Saeedi}, {Santangelo}, {Sasaki}, {Schmitt}, {Schneider}, {Schrabback}, {Schuster}, {Schwope}, {Seppi}, {Serim}, {Shreeram}, {Sokolova-Lapa}, {Starck}, {Stelzer}, {Stierhof}, {Suleimanov}, {Tenzer}, {Traulsen}, {Tr{\"u}mper}, {Tsuge}, {Urrutia}, {Veronica}, {Waddell}, {Willer}, {Wolf}, {Yeung}, {Zainab}, {Zangrandi}, {Zhang}, {Zhang}, \& {Zheng}}]{Merloni2024A&A}
{Merloni}, A., {Lamer}, G., {Liu}, T., {et~al.} 2024{\natexlab{a}}, \aap, 682, A34

\bibitem[{{Merloni} {et~al.}(2024{\natexlab{b}}){Merloni}, {Lamer}, {Liu}, {Ramos-Ceja}, {Brunner}, {Dennerl}, {Doroshenko}, {Freyberg}, {Friedrich}, {Gatuzz}, {Georgakakis}, {Haberl}, {Igo}, {Kreykenbohm}, {Liu}, {Maitra}, {Malyali}, {Mayer}, {Nandra}, {Predehl}, {Robrade}, {Salvato}, {Sanders}, {Stewart}, {Tubin-Arenas}, {Weber}, {Wilms}, {Arcodia}, {Artis}, {Aschersleben}, {Avakyan}, {Aydar}, {Bahar}, {Balzer}, {Becker}, {Berger}, {Boller}, {Bornemann}, {Brueggen}, {Brusa}, {Buchner}, {Burwitz}, {Camilloni}, {Clerc}, {Comparat}, {Coutinho}, {Czesla}, {Dannhauer}, {Dauner}, {Dauser}, {Dietl}, {Dolag}, {Dwelly}, {Egg}, {Ehl}, {Freund}, {Friedrich}, {Gaida}, {Garrel}, {Ghirardini}, {Gokus}, {Gruenwald}, {Grandis}, {Grotova}, {Gruen}, {Gueguen}, {Haemmerich}, {Hamaus}, {Hasinger}, {Haubner}, {Homan}, {Ider Chitham}, {Joseph}, {Joyce}, {Koenig}, {Kaltenbrunner}, {Khokhriakova}, {Kink}, {Kirsch}, {Kluge}, {Knies}, {Krippendorf}, {Krumpe}, {Kurpas}, {Li}, {Liu}, {Locatelli}, {Lorenz}, {Mueller}, {Magaudda},
  {Mannes}, {McCall}, {Meidinger}, {Michailidis}, {Migkas}, {Munoz-Giraldo}, {Musiimenta}, {Nguyen-Dang}, {Ni}, {Olechowska}, {Ota}, {Pacaud}, {Pasini}, {Perinati}, {Pires}, {Pommranz}, {Ponti}, {Poppenhaeger}, {Puehlhofer}, {Rau}, {Reh}, {Reiprich}, {Roster}, {Saeedi}, {Santangelo}, {Sasaki}, {Schmitt}, {Schneider}, {Schrabback}, {Schuster}, {Schwope}, {Seppi}, {Serim}, {Shreeram}, {Sokolova-Lapa}, {Starck}, {Stelzer}, {Stierhof}, {Suleimanov}, {Tenzer}, {Traulsen}, {Truemper}, {Tsuge}, {Urrutia}, {Veronica}, {Waddell}, {Willer}, {Wolf}, {Yeung}, {Zainab}, {Zangrandi}, {Zhang}, {Zhang}, \& {Zheng}}]{Merloni2024yCat}
{Merloni}, A., {Lamer}, G., {Liu}, T., {et~al.} 2024{\natexlab{b}}, {VizieR Online Data Catalog: SRG/eROSITA all-sky survey catalogs (eRASS1) (Merloni+, 2024)}, VizieR On-line Data Catalog: J/A+A/682/A34. Originally published in: 2024A\&A...682A..34M

\bibitem[{{Osterbrock} \& {Ferland}(2006)}]{Osterbrock2006}
{Osterbrock}, D.~E. \& {Ferland}, G.~J. 2006, {Astrophysics of gaseous nebulae and active galactic nuclei}

\bibitem[{{Panessa} {et~al.}(2006){Panessa}, {Bassani}, {Cappi}, {Dadina}, {Barcons}, {Carrera}, {Ho}, \& {Iwasawa}}]{Panessa2006}
{Panessa}, F., {Bassani}, L., {Cappi}, M., {et~al.} 2006, \aap, 455, 173

\bibitem[{{Planck Collaboration} {et~al.}(2020){Planck Collaboration}, {Aghanim}, {Akrami}, {Ashdown}, {Aumont}, {Baccigalupi}, {Ballardini}, {Banday}, {Barreiro}, {Bartolo}, {Basak}, {Battye}, {Benabed}, {Bernard}, {Bersanelli}, {Bielewicz}, {Bock}, {Bond}, {Borrill}, {Bouchet}, {Boulanger}, {Bucher}, {Burigana}, {Butler}, {Calabrese}, {Cardoso}, {Carron}, {Challinor}, {Chiang}, {Chluba}, {Colombo}, {Combet}, {Contreras}, {Crill}, {Cuttaia}, {de Bernardis}, {de Zotti}, {Delabrouille}, {Delouis}, {Di Valentino}, {Diego}, {Dor{\'e}}, {Douspis}, {Ducout}, {Dupac}, {Dusini}, {Efstathiou}, {Elsner}, {En{\ss}lin}, {Eriksen}, {Fantaye}, {Farhang}, {Fergusson}, {Fernandez-Cobos}, {Finelli}, {Forastieri}, {Frailis}, {Fraisse}, {Franceschi}, {Frolov}, {Galeotta}, {Galli}, {Ganga}, {G{\'e}nova-Santos}, {Gerbino}, {Ghosh}, {Gonz{\'a}lez-Nuevo}, {G{\'o}rski}, {Gratton}, {Gruppuso}, {Gudmundsson}, {Hamann}, {Handley}, {Hansen}, {Herranz}, {Hildebrandt}, {Hivon}, {Huang}, {Jaffe}, {Jones}, {Karakci}, {Keih{\"a}nen},
  {Keskitalo}, {Kiiveri}, {Kim}, {Kisner}, {Knox}, {Krachmalnicoff}, {Kunz}, {Kurki-Suonio}, {Lagache}, {Lamarre}, {Lasenby}, {Lattanzi}, {Lawrence}, {Le Jeune}, {Lemos}, {Lesgourgues}, {Levrier}, {Lewis}, {Liguori}, {Lilje}, {Lilley}, {Lindholm}, {L{\'o}pez-Caniego}, {Lubin}, {Ma}, {Mac{\'\i}as-P{\'e}rez}, {Maggio}, {Maino}, {Mandolesi}, {Mangilli}, {Marcos-Caballero}, {Maris}, {Martin}, {Martinelli}, {Mart{\'\i}nez-Gonz{\'a}lez}, {Matarrese}, {Mauri}, {McEwen}, {Meinhold}, {Melchiorri}, {Mennella}, {Migliaccio}, {Millea}, {Mitra}, {Miville-Desch{\^e}nes}, {Molinari}, {Montier}, {Morgante}, {Moss}, {Natoli}, {N{\o}rgaard-Nielsen}, {Pagano}, {Paoletti}, {Partridge}, {Patanchon}, {Peiris}, {Perrotta}, {Pettorino}, {Piacentini}, {Polastri}, {Polenta}, {Puget}, {Rachen}, {Reinecke}, {Remazeilles}, {Renzi}, {Rocha}, {Rosset}, {Roudier}, {Rubi{\~n}o-Mart{\'\i}n}, {Ruiz-Granados}, {Salvati}, {Sandri}, {Savelainen}, {Scott}, {Shellard}, {Sirignano}, {Sirri}, {Spencer}, {Sunyaev}, {Suur-Uski}, {Tauber}, {Tavagnacco},
  {Tenti}, {Toffolatti}, {Tomasi}, {Trombetti}, {Valenziano}, {Valiviita}, {Van Tent}, {Vibert}, {Vielva}, {Villa}, {Vittorio}, {Wandelt}, {Wehus}, {White}, {White}, {Zacchei}, \& {Zonca}}]{Planck2018}
{Planck Collaboration}, {Aghanim}, N., {Akrami}, Y., {et~al.} 2020, \aap, 641, A6

\bibitem[{{Predehl} {et~al.}(2021){Predehl}, {Andritschke}, {Arefiev}, {Babyshkin}, {Batanov}, {Becker}, {B{\"o}hringer}, {Bogomolov}, {Boller}, {Borm}, {Bornemann}, {Br{\"a}uninger}, {Br{\"u}ggen}, {Brunner}, {Brusa}, {Bulbul}, {Buntov}, {Burwitz}, {Burkert}, {Clerc}, {Churazov}, {Coutinho}, {Dauser}, {Dennerl}, {Doroshenko}, {Eder}, {Emberger}, {Eraerds}, {Finoguenov}, {Freyberg}, {Friedrich}, {Friedrich}, {F{\"u}rmetz}, {Georgakakis}, {Gilfanov}, {Granato}, {Grossberger}, {Gueguen}, {Gureev}, {Haberl}, {H{\"a}lker}, {Hartner}, {Hasinger}, {Huber}, {Ji}, {Kienlin}, {Kink}, {Korotkov}, {Kreykenbohm}, {Lamer}, {Lomakin}, {Lapshov}, {Liu}, {Maitra}, {Meidinger}, {Menz}, {Merloni}, {Mernik}, {Mican}, {Mohr}, {M{\"u}ller}, {Nandra}, {Nazarov}, {Pacaud}, {Pavlinsky}, {Perinati}, {Pfeffermann}, {Pietschner}, {Ramos-Ceja}, {Rau}, {Reiffers}, {Reiprich}, {Robrade}, {Salvato}, {Sanders}, {Santangelo}, {Sasaki}, {Scheuerle}, {Schmid}, {Schmitt}, {Schwope}, {Shirshakov}, {Steinmetz}, {Stewart}, {Str{\"u}der},
  {Sunyaev}, {Tenzer}, {Tiedemann}, {Tr{\"u}mper}, {Voron}, {Weber}, {Wilms}, \& {Yaroshenko}}]{Predehl2021}
{Predehl}, P., {Andritschke}, R., {Arefiev}, V., {et~al.} 2021, \aap, 647, A1

\bibitem[{{Pucha} {et~al.}(2024){Pucha}, {Juneau}, {Dey}, {Siudek}, {Mezcua}, {Moustakas}, {BenZvi}, {Hainline}, {Hviding}, {Mao}, {Alexander}, {Alfarsy}, {Circosta}, {Guo}, {Manwadkar}, {Martini}, {Weaver}, {Aguilar}, {Ahlen}, {Bianchi}, {Brooks}, {Canning}, {Claybaugh}, {Dawson}, {de la Macorra}, {Dey}, {Doel}, {Font-Ribera}, {Forero-Romero}, {Gazta{\~n}aga}, {Gontcho}, {Gutierrez}, {Honscheid}, {Kehoe}, {Koposov}, {Lambert}, {Landriau}, {Le Guillou}, {Meisner}, {Miquel}, {Prada}, {Rossi}, {Sanchez}, {Schlegel}, {Schubnell}, {Seo}, {Sprayberry}, {Tarl{\'e}}, \& {Zou}}]{Pucha2024}
{Pucha}, R., {Juneau}, S., {Dey}, A., {et~al.} 2024, arXiv e-prints, arXiv:2411.00091

\bibitem[{{Pulatova} {et~al.}(2024){Pulatova}, {Rix}, {Tugay}, {Zadorozhna}, {Seeburger}, \& {Demianenko}}]{Pulatova2024}
{Pulatova}, N.~G., {Rix}, H.~W., {Tugay}, A.~V., {et~al.} 2024, \aap, 686, A223

\bibitem[{{Pulatova} {et~al.}(2015){Pulatova}, {Vavilova}, {Sawangwit}, {Babyk}, \& {Klimanov}}]{Pulatova2015}
{Pulatova}, N.~G., {Vavilova}, I.~B., {Sawangwit}, U., {Babyk}, I., \& {Klimanov}, S. 2015, \mnras, 447, 2209

\bibitem[{{Reines} {et~al.}(2013){Reines}, {Greene}, \& {Geha}}]{Reines2013}
{Reines}, A.~E., {Greene}, J.~E., \& {Geha}, M. 2013, \apj, 775, 116

\bibitem[{{Rubtsov} {et~al.}(2024{\natexlab{a}}){Rubtsov}, {Chilingarian}, {Katkov}, {Grishin}, {Goradzhanov}, \& {Borisov}}]{2024ASPC..535..371R}
{Rubtsov}, E., {Chilingarian}, I., {Katkov}, I., {et~al.} 2024{\natexlab{a}}, in Astronomical Society of the Pacific Conference Series, Vol. 535, Astromical Data Analysis Software and Systems XXXI, ed. B.~V. {Hugo}, R.~{Van Rooyen}, \& O.~M. {Smirnov}, 371

\bibitem[{{Rubtsov} {et~al.}(2024{\natexlab{b}}){Rubtsov}, {Chilingarian}, {Katkov}, {Grishin}, {Kasparova}, {Toptun}, \& {Goradzhanov}}]{2024IAUGA..32P2683R}
{Rubtsov}, E., {Chilingarian}, I., {Katkov}, I., {et~al.} 2024{\natexlab{b}}, in IAU General Assembly, 2683

\bibitem[{{Ryzhov} {et~al.}(2024){Ryzhov}, {Micha{\l}owski}, {Nadolny}, {Hjorth}, {Le{\'s}niewska}, {Solar}, {Nowaczyk}, {Gall}, \& {Takeuchi}}]{Ryzhov2024}
{Ryzhov}, O., {Micha{\l}owski}, M.~J., {Nadolny}, J., {et~al.} 2024, arXiv e-prints, arXiv:2411.10517

\bibitem[{{Salvato} {et~al.}(2018){Salvato}, {Buchner}, {Budav{\'a}ri}, {Dwelly}, {Merloni}, {Brusa}, {Rau}, {Fotopoulou}, \& {Nandra}}]{Salvato2018}
{Salvato}, M., {Buchner}, J., {Budav{\'a}ri}, T., {et~al.} 2018, \mnras, 473, 4937

\bibitem[{{Salvato} {et~al.}(2022){Salvato}, {Wolf}, {Dwelly}, {Georgakakis}, {Brusa}, {Merloni}, {Liu}, {Toba}, {Nandra}, {Lamer}, {Buchner}, {Schneider}, {Freund}, {Rau}, {Schwope}, {Nishizawa}, {Klein}, {Arcodia}, {Comparat}, {Musiimenta}, {Nagao}, {Brunner}, {Malyali}, {Finoguenov}, {Anderson}, {Shen}, {Ibarra-Medel}, {Trump}, {Brandt}, {Urry}, {Rivera}, {Krumpe}, {Urrutia}, {Miyaji}, {Ichikawa}, {Schneider}, {Fresco}, {Boller}, {Haase}, {Brownstein}, {Lane}, {Bizyaev}, \& {Nitschelm}}]{Salvato2022}
{Salvato}, M., {Wolf}, J., {Dwelly}, T., {et~al.} 2022, \aap, 661, A3

\bibitem[{Salvato \& Wolf(2025)}]{Salvato2025}
Salvato, M. \& Wolf, J. e.~a. 2025, Counterpart identification and classification for eRASS1 and characterisation of the AGN content, submitted

\bibitem[{{Shapiro} {et~al.}(2009){Shapiro}, {Genzel}, {Quataert}, {F{\"o}rster Schreiber}, {Davies}, {Tacconi}, {Armus}, {Bouch{\'e}}, {Buschkamp}, {Cimatti}, {Cresci}, {Daddi}, {Eisenhauer}, {Erb}, {Genel}, {Hicks}, {Lilly}, {Lutz}, {Renzini}, {Shapley}, {Steidel}, \& {Sternberg}}]{2009ApJ...701..955S}
{Shapiro}, K.~L., {Genzel}, R., {Quataert}, E., {et~al.} 2009, \apj, 701, 955

\bibitem[{Skilling(2004)}]{10.1063/1.1835238}
Skilling, J. 2004, AIP Conference Proceedings, 735, 395

\bibitem[{{Sunyaev} {et~al.}(2021){Sunyaev}, {Arefiev}, {Babyshkin}, {Bogomolov}, {Borisov}, {Buntov}, {Brunner}, {Burenin}, {Churazov}, {Coutinho}, {Eder}, {Eismont}, {Freyberg}, {Gilfanov}, {Gureyev}, {Hasinger}, {Khabibullin}, {Kolmykov}, {Komovkin}, {Krivonos}, {Lapshov}, {Levin}, {Lomakin}, {Lutovinov}, {Medvedev}, {Merloni}, {Mernik}, {Mikhailov}, {Molodtsov}, {Mzhelsky}, {M{\"u}ller}, {Nandra}, {Nazarov}, {Pavlinsky}, {Poghodin}, {Predehl}, {Robrade}, {Sazonov}, {Scheuerle}, {Shirshakov}, {Tkachenko}, \& {Voron}}]{Sunyaev2021}
{Sunyaev}, R., {Arefiev}, V., {Babyshkin}, V., {et~al.} 2021, \aap, 656, A132

\bibitem[{{Taylor}(2005)}]{Taylor2005}
{Taylor}, M.~B. 2005, in Astronomical Data Analysis Software and Systems XIV, Vol. 347, 29

\bibitem[{{Tremaine} {et~al.}(2002){Tremaine}, {Gebhardt}, {Bender}, {Bower}, {Dressler}, {Faber}, {Filippenko}, {Green}, {Grillmair}, {Ho}, {Kormendy}, {Lauer}, {Magorrian}, {Pinkney}, \& {Richstone}}]{Tremaine2002}
{Tremaine}, S., {Gebhardt}, K., {Bender}, R., {et~al.} 2002, \apj, 574, 740

\bibitem[{Ueda {et~al.}(2015)Ueda, Hashimoto, Ichikawa, Ishino, Kniazev, Väisänen, Ricci, Berney, Gandhi, Koss, Mushotzky, Terashima, Trakhtenbrot, \& Crenshaw}]{Ueda2015}
Ueda, Y., Hashimoto, Y., Ichikawa, K., {et~al.} 2015, The Astrophysical Journal, 815, 1

\bibitem[{{Urry} \& {Padovani}(1995)}]{Urry1995}
{Urry}, C.~M. \& {Padovani}, P. 1995, \pasp, 107, 803

\bibitem[{{van der Marel} \& {Franx}(1993)}]{1993ApJ...407..525V}
{van der Marel}, R.~P. \& {Franx}, M. 1993, \apj, 407, 525

\bibitem[{{Vazdekis} {et~al.}(2016){Vazdekis}, {Koleva}, {Ricciardelli}, {R{\"o}ck}, \& {Falc{\'o}n-Barroso}}]{2016MNRAS.463.3409V}
{Vazdekis}, A., {Koleva}, M., {Ricciardelli}, E., {R{\"o}ck}, B., \& {Falc{\'o}n-Barroso}, J. 2016, \mnras, 463, 3409

\bibitem[{{Veilleux} \& {Osterbrock}(1987)}]{Veilleux1987}
{Veilleux}, S. \& {Osterbrock}, D.~E. 1987, \apjs, 63, 295

\bibitem[{{Wang} {et~al.}(2024){Wang}, {Woo}, {Gallo}, {Guo}, {Son}, {Kong}, {Mandal}, {Cho}, {Kim}, \& {Shin}}]{Wang2024}
{Wang}, S., {Woo}, J.-H., {Gallo}, E., {et~al.} 2024, \apj, 966, 128

\bibitem[{{Yang} {et~al.}(2020){Yang}, {Boquien}, {Buat}, {Burgarella}, {Ciesla}, {Duras}, {Stalevski}, {Brandt}, \& {Papovich}}]{Yang2020}
{Yang}, G., {Boquien}, M., {Buat}, V., {et~al.} 2020, \mnras, 491, 740

\bibitem[{{York} {et~al.}(2000){York}, {Adelman}, {Anderson}, {Anderson}, {Annis}, {Bahcall}, {Bakken}, {Barkhouser}, {Bastian}, {Berman}, {Boroski}, {Bracker}, {Briegel}, {Briggs}, {Brinkmann}, {Brunner}, {Burles}, {Carey}, {Carr}, {Castander}, {Chen}, {Colestock}, {Connolly}, {Crocker}, {Csabai}, {Czarapata}, {Davis}, {Doi}, {Dombeck}, {Eisenstein}, {Ellman}, {Elms}, {Evans}, {Fan}, {Federwitz}, {Fiscelli}, {Friedman}, {Frieman}, {Fukugita}, {Gillespie}, {Gunn}, {Gurbani}, {de Haas}, {Haldeman}, {Harris}, {Hayes}, {Heckman}, {Hennessy}, {Hindsley}, {Holm}, {Holmgren}, {Huang}, {Hull}, {Husby}, {Ichikawa}, {Ichikawa}, {Ivezi{\'c}}, {Kent}, {Kim}, {Kinney}, {Klaene}, {Kleinman}, {Kleinman}, {Knapp}, {Korienek}, {Kron}, {Kunszt}, {Lamb}, {Lee}, {Leger}, {Limmongkol}, {Lindenmeyer}, {Long}, {Loomis}, {Loveday}, {Lucinio}, {Lupton}, {MacKinnon}, {Mannery}, {Mantsch}, {Margon}, {McGehee}, {McKay}, {Meiksin}, {Merelli}, {Monet}, {Munn}, {Narayanan}, {Nash}, {Neilsen}, {Neswold}, {Newberg}, {Nichol}, {Nicinski},
  {Nonino}, {Okada}, {Okamura}, {Ostriker}, {Owen}, {Pauls}, {Peoples}, {Peterson}, {Petravick}, {Pier}, {Pope}, {Pordes}, {Prosapio}, {Rechenmacher}, {Quinn}, {Richards}, {Richmond}, {Rivetta}, {Rockosi}, {Ruthmansdorfer}, {Sandford}, {Schlegel}, {Schneider}, {Sekiguchi}, {Sergey}, {Shimasaku}, {Siegmund}, {Smee}, {Smith}, {Snedden}, {Stone}, {Stoughton}, {Strauss}, {Stubbs}, {SubbaRao}, {Szalay}, {Szapudi}, {Szokoly}, {Thakar}, {Tremonti}, {Tucker}, {Uomoto}, {Vanden Berk}, {Vogeley}, {Waddell}, {Wang}, {Watanabe}, {Weinberg}, {Yanny}, {Yasuda}, \& {SDSS Collaboration}}]{York2000}
{York}, D.~G., {Adelman}, J., {Anderson}, Jr., J.~E., {et~al.} 2000, \aj, 120, 1579

\end{thebibliography}

\begin{appendix}

\onecolumn

\section{Description of the table columns}
\label{app:tab}

\begin{strip}
\begin{center}
\begin{longtable}{llll}
\caption{Table description of optical emission line properties of eROSITA-selected SDSS-V galaxies}
\\ \hline \hline
Column Name & Units & Datatype & Description \\
\hline \hline
\endfirsthead
\caption{Table description of optical emission line properties of eROSITA-selected SDSS-V galaxies (cont.)}
\\ \hline \hline
Column Name & Units & Datatype & Description \\
\hline \hline
\endhead
\hline \multicolumn{4}{r}{{Continued on next page}} \\
\endfoot
\hline
\endlastfoot
plateid &   & long64 & SDSS-V plate ID \\
mjd & days & long64 & Modified Julian Date of the observation \\
catalogid &   & long64 & SDSS-V catalog ID for target \\
sdssid &   & long64 & Unique photometric identifier in the SDSS database \\
specobjid &   & string & Unique spectrum identifier in the SDSS database \\
erass1 &   & string & String containing the official IAU name of the eRASS1 source \\
ra & deg & double & Right ascension (ICRS) for SDSS-V target \\
dec & deg & double & Declination (ICRS) for SDSS-V target \\
xra & deg & double & Right ascension (ICRS) from eRASS1 \\
xdec & deg & double & Declination (ICRS) from eRASS1 \\
distance & arcsec & double & Distance between SDSS-V and eRASS1 positions \\
z &   & double & Redshift \\
z\_e &   & double & Redshift uncertainty \\
z\_q &   & integer & Redshift flag (ZWARNING from SDSS-V) \\
quality &   & integer & Quality flag (-1/0/1/2) see Sec.~\ref{subsec:table} \\
mag\_u & mag & float & Magnitude in SDSS u filter \\
mag\_g & mag & float & Magnitude in SDSS g filter \\
mag\_r & mag & float & Magnitude in SDSS r filter \\
mag\_i & mag & float & Magnitude in SDSS i filter \\
mag\_z & mag & float & Magnitude in SDSS z filter \\
err\_u & mag & float & Magnitude uncertainty in SDSS u filter \\
err\_g & mag & float & Magnitude uncertainty in SDSS g filter \\
err\_r & mag & float & Magnitude uncertainty in SDSS r filter \\
err\_i & mag & float & Magnitude uncertainty in SDSS i filter \\
err\_z & mag & float & Magnitude uncertainty in SDSS z filter \\
snr\_flux &   & float & Median SNR of the SDSS-V spectrum \\
snr\_fit &   & float & Median SNR of the best-fit model \\
snr\_grid &   & float & Median SNR of the stellar component (E-MILES SSP model) \\
snr\_cont &   & float & Median SNR at the continuum level (excluded emission lines) \\
chisqr &   & float & Reduces $\chi^2$ statistics for {\sc NBursts} fit \\
dof &   & integer & Degree of freedom for {\sc NBursts} fit \\
vel\_ssp & km s$^{-1}$ & float & Radial velocity of stellar component \\
vel\_nlr & km s$^{-1}$ & float & Radial velocity of NLR component \\
vel\_blr & km s$^{-1}$ & float & Radial velocity of BLR component \\
sig\_ssp & km s$^{-1}$ & float & Velocity dispersion of stellar component \\
sig\_nlr & km s$^{-1}$ & float & Velocity dispersion of NLR component \\
sig\_blr & km s$^{-1}$ & float & Velocity dispersion of BLR component \\
h3\_ssp &   & float & $h_3$ Gauss-Hermite coefficient of stellar component \\
h3\_nlr &   & float & $h_3$ Gauss-Hermite coefficient of NLR component \\
h3\_blr &   & float & $h_3$ Gauss-Hermite coefficient of BLR component \\
h4\_ssp &   & float & $h_4$ Gauss-Hermite coefficient of stellar component \\
h4\_nlr &   & float & $h_4$ Gauss-Hermite coefficient of NLR component \\
h4\_blr &   & float & $h_4$ Gauss-Hermite coefficient of BLR component \\
age & Myr & float & Age of stellar component (E-MILES SSP model) \\
met & dex & float & Metallicity of stellar component (E-MILES SSP model) \\
e\_vel\_ssp & km s$^{-1}$ & float & Radial velocity uncertainty of stellar component \\
e\_vel\_nlr & km s$^{-1}$ & float & Radial velocity uncertainty of NLR component \\
e\_vel\_blr & km s$^{-1}$ & float & Radial velocity uncertainty of BLR component \\
e\_sig\_ssp & km s$^{-1}$ & float & Velocity dispersion uncertainty of stellar component \\
e\_sig\_nlr & km s$^{-1}$ & float & Velocity dispersion uncertainty of NLR component \\
e\_sig\_blr & km s$^{-1}$ & float & Velocity dispersion uncertainty of BLR component \\
e\_h3\_ssp &   & float & $h_3$ Gauss-Hermite coefficient uncertainty of stellar component \\
e\_h3\_nlr &   & float & $h_3$ Gauss-Hermite coefficient uncertainty of NLR component \\
e\_h3\_blr &   & float & $h_3$ Gauss-Hermite coefficient uncertainty of BLR component \\
e\_h4\_ssp &   & float & $h_4$ Gauss-Hermite coefficient uncertainty of stellar component \\
e\_h4\_nlr &   & float & $h_4$ Gauss-Hermite coefficient uncertainty of NLR component \\
e\_h4\_blr &   & float & $h_4$ Gauss-Hermite coefficient uncertainty of BLR component \\
e\_age & Myr & float & Age uncertainty of stellar component (E-MILES SSP model) \\
e\_met & dex & float & Metallicity uncertainty of stellar component (E-MILES SSP model) \\
ebv & mag & float & Intrinsic E(B-V) \\
e\_ebv & mag & float & Uncertainty of intrinsic E(B-V) \\
mbh & $M_{\odot}$ & float & Black hole mass \citep{Reines2013} \\
e\_mbh & $M_{\odot}$ & float & Black hole mass uncertainty \\
f\_oii3727\_nlr & erg s$^{-1}$ cm$^{-2}$ & float & Emission line flux (example for [O\ii]~$\lambda 3727$, see full emission lines list at App~\ref{app:eml}) \\
e\_oii3727\_nlr & erg s$^{-1}$ cm$^{-2}$ & float & Emission line flux uncertainty \\
c\_oii3727\_nlr & erg s$^{-1}$ cm$^{-2} \AA^{-1}$ & float & Continuum level at emission line wavelength \\
ml\_flux\_1 & erg s$^{-1}$ cm$^{-2} \AA^{-1}$ & float & Source X-ray flux (example for ``1'', see full eRASS1 energy bands list at App~\ref{app:xray}) \\
ml\_flux\_err\_1 & erg s$^{-1}$ cm$^{-2} \AA^{-1}$ & float & Source X-ray flux uncertainty \\
\hline
\end{longtable}
\end{center}
\twocolumn
\end{strip}

\section{Optical emission lines list.}
\label{app:eml}

\begin{table}[h]
\centering
\begin{tabular}{llcc}
\hline
\hline
& & Vacuum & Air \\
Line ID & Label & Wavelength,  & Wavelength, \\
& & $\AA$ & $\AA$ \\
\hline
\hline
oii3727 & [O \ii] & 3727.092 & 3726.032 \\
oii3729 & [O \ii] & 3729.875 & 3728.815 \\
neiii3869 & [Ne \iii] & 3869.857 & 3868.760 \\
h83890$^{b}$ & H8 & 3890.166 & 3889.064 \\
hei3965 & He \i & 3965.850 & 3964.728 \\
neiii3968 & [Ne \iii] & 3968.593 & 3967.470 \\
hepsilon$^{b}$ & H$\epsilon$ & 3971.202 & 3970.079 \\
hdelta$^{b}$ & H$\delta$ & 4102.900 & 4101.742 \\
hgamma$^{b}$ & H$\gamma$ & 4341.692 & 4340.472 \\
oiii4364 & [O \iii] & 4364.437 & 4363.211 \\
hei4389 & He \i & 4389.160 & 4387.927 \\
oii4652 & O \ii & 4652.140 & 4650.838 \\
heii4687 & He \ii & 4687.022 & 4685.711 \\
hei4714 & He \i & 4714.470 & 4713.151 \\
hbeta$^{b}$ & H$\beta$ & 4862.692 & 4861.334 \\
hei4923 & He \i & 4923.310 & 4921.936 \\
oiii4960 & [O \iii] & 4960.296 & 4958.912 \\
oiii5008 & [O \iii] & 5008.241 & 5006.844 \\
hei5017 & He \i & 5017.080 & 5015.681 \\
hei5877 & He \i & 5877.255 & 5875.626 \\
oi6302 & [O \i] & 6302.049 & 6300.307 \\
oi6365 & [O \i] & 6365.538 & 6363.779 \\
nii6549 & [N \ii] & 6549.862 & 6548.053 \\
halpha$^{b}$ & H$\alpha$ & 6564.635 & 6562.822 \\
nii6585 & [N\ii] & 6585.282 & 6583.463 \\
hei6680 & He \i & 6679.995 & 6678.151 \\
sii6718 & [S \ii] & 6718.298 & 6716.444 \\
sii6732 & [S \ii] & 6732.671 & 6730.813 \\
\hline
\end{tabular}
\caption{List of the main optical emission lines in the rest frame wavelength range $3700 - 7000 \AA$, which we used in the fitting procedure (28 narrow and 6 broad emission lines, $^{b}$ indicates both narrow and broad lines, otherwise narrow line only).}
\end{table}

\pagebreak

\section{List of eRASS1 energy bands.}
\label{app:xray}
\begin{table}[h]
\centering
\begin{tabular}{cc}
\hline
\hline
Band & Energy range, keV \\
\hline
\hline
1 & $0.2 - 2.3$ \\
P1 & $0.2 - 0.5$ \\
P2 & $0.5 - 1.0$ \\
P3 & $1.0 - 2.0$ \\
P4 & $2.0 - 5.0$ \\
P5 & $5.0 - 8.0$ \\
P6 & $4.0 - 10.0$ \\
P7 & $5.1 - 6.1$ \\
P8 & $6.2 - 7.1$ \\
P9 & $7.2 - 8.2$ \\
\hline
\end{tabular}
\caption{List of energy band suffixes in the eRASS1 catalog \citep{Merloni2024yCat, Merloni2024A&A}}
\end{table}

\end{appendix}

\end{document}